# A General Principle: $\pi$-Conjugated Confinement Maximizes Band Gap of DUV NLO Materials

Lin Xiong, Li-Ming Wu,* and Ling Chen*

[*] L. Xiong, Prof. L.-M. Wu
Beijing Key Laboratory of Energy Conversion and Storage Materials
Ministry of Education, College of Chemistry
Beijing Normal University, Beijing 100875 (P. R. China)
E-mail: wlm@bnu.edu.cn ORCID Li-Ming Wu: 0000-0001-8390-2138

Prof. Dr. L. Chen
Key Laboratory of Theoretical and Computational Photochemistry
College of Chemistry, Beijing Normal University
Beijing 100875 (P. R. China)
E-mail: chenl@bnu.edu.cn ORCID Ling Chen: 0000-0002-3693-4193

Supporting information for this article is given via a link at the end of the document.

**Abstract:** Current nonlinear optical materials face a conventional limitation on the tradeoff between band gap and birefringence, especially in the deep UV spectral region. To circumvent such a dilemma, we propose a general principle, a $\pi$-conjugated confinement, to partially decouple the inter group $\pi$-conjugated interactions with the separation of a non-$\pi$-conjugated group so as to maximize the band gap in comparison with those of simple $\pi$-conjugated salts, such as borates, carbonates. Meanwhile, to maintain a large optical anisotropy. We uncover that the $\pi$-conjugated confinement is a shared structural feature for all the known DUV NLO materials with favorable properties (45 compounds), and thus, it provides an essential design criterion. Guided by this principle, the carbonophosphate is predicted theoretically for the first time as a promising DUV candidate system, $Sr_3Y[PO_4][CO_3]_3$ and $Na_3X[PO_4][CO_3]$ (X = Ba, Sr, Ca, Mg) exhibit an enhanced birefringence that is 3–24 times larger than that of the simple phosphate, as well as an increased band gap that is 0.2–1.7 eV wider than that of the simple carbonate. Especially, the shortest SHG output of $Sr_3Y[PO_4][CO_3]_3$ is at $\lambda_{PM}$ = 181 nm, being the shortest one among phosphates to date.

## Introduction

Deep-ultraviolet (DUV) nonlinear optical (NLO) materials playing a crucial role in many livelihood engineering and industrial fields, such as precise detection, laser micro-matching, manufacturing and medical technology, are facing a great challenge that lies between urgent demands and extreme material scarcity.[1] The sole commercial material, $KBe_2BO_3F_2$ (KBBF), possesses a phase-matching second harmonic generation (SHG) output wavelength ($\lambda_{PM}$ ~161 nm)[2] generated by the direct SHG, unfortunately suffers not only a fatal beryllium component but also an undesired strong layered-crystal-growth-habit that brings severe application problems.[1] As the key criteria for a DUV NLO material, the $\lambda_{PM}$ is associated with, first, a wide band gap ($E_g$ > 6.2 eV, i.e., an absorption edge ($\lambda_{cutoff}$) shorter than 200 nm); and secondly, a sufficiently large birefringence ($\Delta n$).[3]

The known structure building unit for all NLO materials, is classified into two categories, the $\pi$-conjugated-, and non-$\pi$-conjugated groups. The $\pi$-conjugated building unit, such as $[BO_3]^{3-}$ and $[CO_3]^{2-}$, constructs compounds showing large second-order susceptibility ($\chi^{(2)}$) and large $\Delta n$, which are failed to exhibit large $E_g$. The intrinsic reason can be understood with the aid of the Molecular Orbital Theory. For instance, in a simple hetero atom system, the splitting energy is written as $\Delta E = \frac{H_{22}+H_{12}}{1-S_{12}} - \frac{H_{11}+H_{12}}{1+S_{12}}$, where the $H_{ii}$ means the Coulomb integral; $H_{ij}$, the exchange integral; and $S_{ij}$, the overlap integral. The $\pi$ bonding interaction, weaker than a $\sigma$ bond interaction, gives rise to a smaller energy splitting $\Delta E$, i.e., the $\pi$-conjugated group exhibits a smaller HOMO–LUMO gap. Whereas, the non $\pi$-conjugated group, characterized by a stronger $\sigma$ bond overlapping, shows a larger HOMO–LUMO gap.

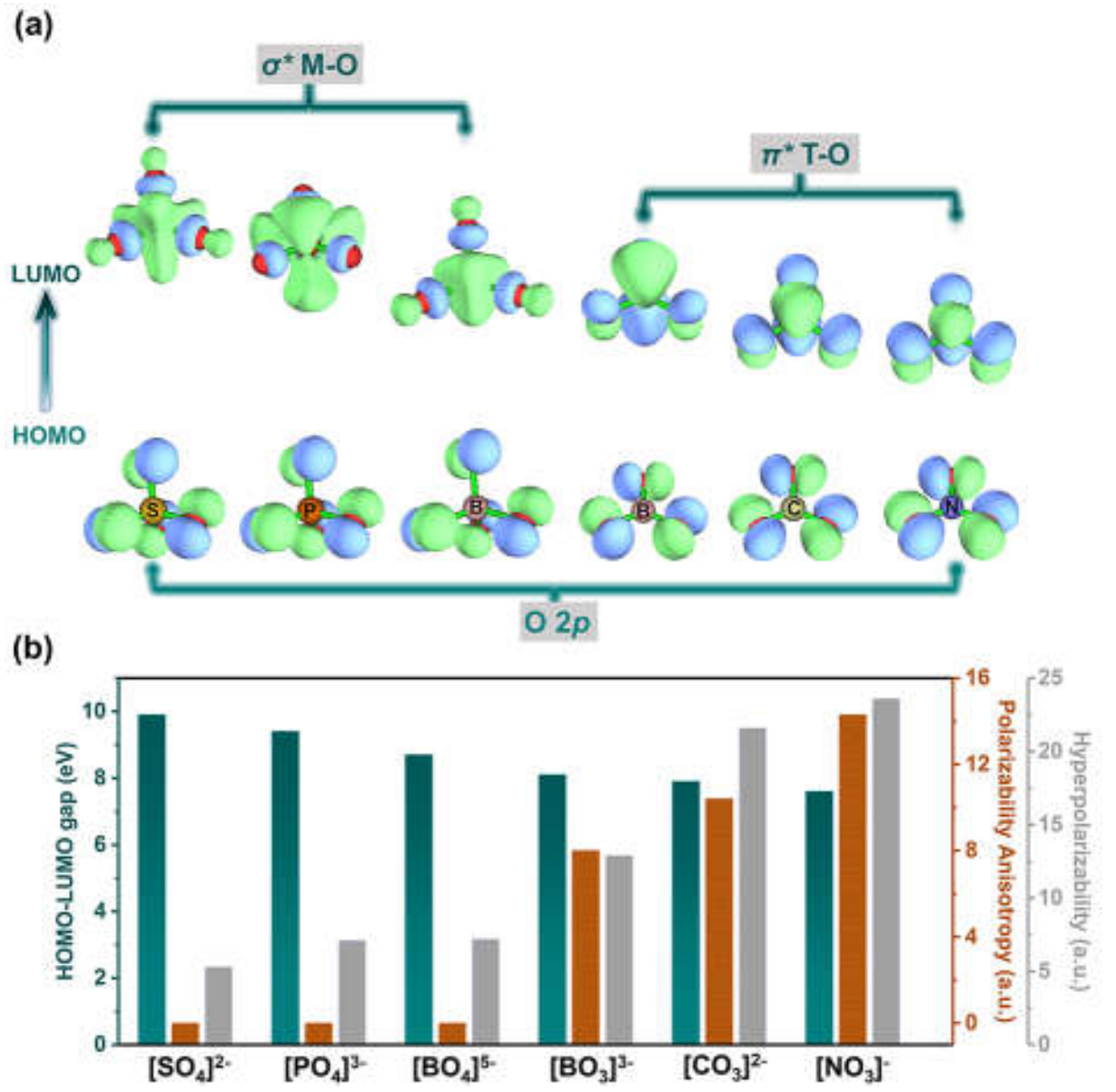

**Figure 1**. (a) HOMO–LUMO molecular orbital distributions and (b) HOMO–LUMO gap, polarizability anisotropy and hyperpolarizability for the tetrahedral $[MO_4]^{x-}$: $[SO_4]^{2-}$, $[PO_4]^{3-}$, $[BO_4]^{5-}$, and planar trigonal $[TO_3]^{x-}$: $[BO_3]^{3-}$, $[CO_3]^{2-}$, $[NO_3]^{-}$ groups.

For instance, having a relatively large $\Delta n$ = 0.119 at 546.1 nm, the famous $\pi$-conjugated borate $\beta$-$BaB_2O_4$ (BBO) exhibits a long $\lambda_{PM}$ of 205 nm that is constrained by the relatively small $E_g$ ($E_g$ = 6.56 eV, i.e., absorption edge ($\lambda_{cutoff}$) = 189 nm),[4] thus, BBO is not a DUV NLO material. For most of the carbonates, their absorption edges are even longer, falling into the range of 200–350 nm.[5] Therefore, carbonate is not a DUV NLO candidate either.

On the other hand, the non-$\pi$-conjugated groups, e.g. tetrahedral $[BO_4]^{5-}$ and $[PO_4]^{3-}$, have exceptional large HOMO–LUMO gaps. The corresponding compounds show very short absorption edges, e.g., $\lambda_{cutoff}$ = 174 or 130 nm to $KH_2PO_4$ (KDP)[6] or $BPO_4$ (BPO).[7] However, constrained by the optical isotropic nature of its non-$\pi$-conjugated $[PO_4]^{3-}$ structural building unit, KDP exhibits small $\Delta n$ (< 0.04 at 1064 nm) that limits the phase matching SHG output wavelength at 259 nm. Therefore, KDP is not used in the higher energy DUV region below 200 nm.[6–8] Nevertheless, among phosphates, the $\Delta n$ of KDP is obviously large, because its building unit $[PO_2(OH)_2]^-$, is considerably squashed from the ideal tetrahedral $[PO_4]^{3-}$ owning to the presence of two hydroxide $[OH]^-$ ligands. And such a geometry distortion has already enlarged the $\Delta n$. A different approach to increase the $[PO_4]^{3-}$ distortion is to utilize a hetero coordination atom as found in a $[PO_3F]^{2-}$.[9] Continuous efforts have realized the nearly perfect alignment of $[PO_3F]^{2-}$ that leads to a great enhancement of $\Delta n$ to 0.053,[1,9] being the largest in the phosphate and fluorophosphate families. Obviously, further enhancement in $\Delta n$ via such a path reaches its up-limit.

In short, the tradeoff between band gap and birefringence roots in such inherent natures of the structure building groups, which become sharp in the upmost high energy DUV spectral region. Therefore, design and searching of high performance DUV NLO materials is challenging and significant.

Herein, we propose a general principle to circumvent the conventional limitation on the tradeoff between band gap and birefringence. We propose a $\pi$-conjugated confinement to decrease the inter-unit $\pi$-conjugated interactions through the separation of the non-$\pi$-conjugated groups so as to maximize $E_g$ in comparison with those of simple $\pi$-conjugated salts. And meanwhile, to keep a suitable density of the $\pi$-conjugated group to maintain a large optical anisotropy. This principle is valid for all the known 45 DUV NLO materials with favorable properties to date. Guided by this principle, we predict the carbonophosphate as a promising DUV NLO candidate system for the first time. $Sr_3Y[PO_4][CO_3]_3$ (**SYPC₃**) and $Na_3X[PO_4][CO_3]$ (**NXPC**, X = Ba, Sr, Ca, Mg), exhibit not only a greatly enhanced $\Delta n$ that is 3–24 times larger than that of simple phosphate, but also an enlarged $E_g$ that is 0.2–1.7 eV wider than that of the simple carbonate. More importantly, **SYPC$_3$** exhibits a $\lambda_{PM}$ = 181 nm, being the shortest SHG output wavelength among phosphates to date. The future experimental works on this and other possible combinations of all the conventional $\pi$-conjugated $[TO_3]^{x-}$ (T = B, C, N) and non-$\pi$-conjugated $[MO_4]^{x-}$ building units are in great anticipation.

## Results and Discussion

### Conventional $\pi$-conjugated and non-$\pi$-conjugated groups show inversely related HOMO–LUMO gap and polarizability anisotropy

We firstly carried out systematic calculations on the conventional $\pi$-conjugated planar $[TO_3]^{x-}$ (T = B, C, N) and non-$\pi$-conjugated tetrahedral $[MO_4]^{x-}$ (M = B, P, S) groups. The results indicate they show similar HOMO orbital spatial distributions that majorly constitute the nonbonding $p$ orbitals coming from O atoms. But their LUMOs are rather different that are contributed either from the $\sigma^*$ M-O bonds, or from the $\pi^*$ T-O bonds. (**Figure 1**) As discussed above, the energy level is generally higher in the $\sigma^*$ M-O bond than in the $\pi^*$ orbital, and the relatively weak $\pi$-$\pi$ conjugated T-O interactions create a smaller energy splitting $\Delta E$. Ongoing from $[BO_3]^{3-}$, $[CO_3]^{2-}$ to $[NO_3]^-$, the $\Delta E$ decreases as the electronegativity difference between T and O atoms declines. And the $\Delta E$ also decreases as the shortening of the T-O bond length. Associated with such a $\Delta E$ decrease, both the polarizability anisotropy ($\delta$) and hyperpolarizability ($|\beta_{max}|$) increase. (**Figure 1**) In the following sections, we only consider the $\pi$-conjugated $[BO_3]^{3-}$ and $[CO_3]^{2-}$, with the exclusion of $[NO_3]^-$, because it is deliquescent. The $\pi_4^6$-$[CO_3]^{2-}$ is isoelectric with the $\pi_4^6$-$[BO_3]^{3-}$, as the length of the C-O bond is shorter than the B-O bond, the $\pi$-bond strength is stronger in $[CO_3]^{2-}$, which is reflected by the facts that its $\beta_{max}$ and $\delta$ are 1.7, and 1.3 times larger than those of $[BO_3]^{3-}$. Regarding SHG intensity and birefringence, carbonate is superior to borate, however, the smaller $E_g$ constrains carbonate from being transparent in the spectral region below 200 nm, and consequently, carbonate cannot satisfy the prerequisite of a DUV NLO material.

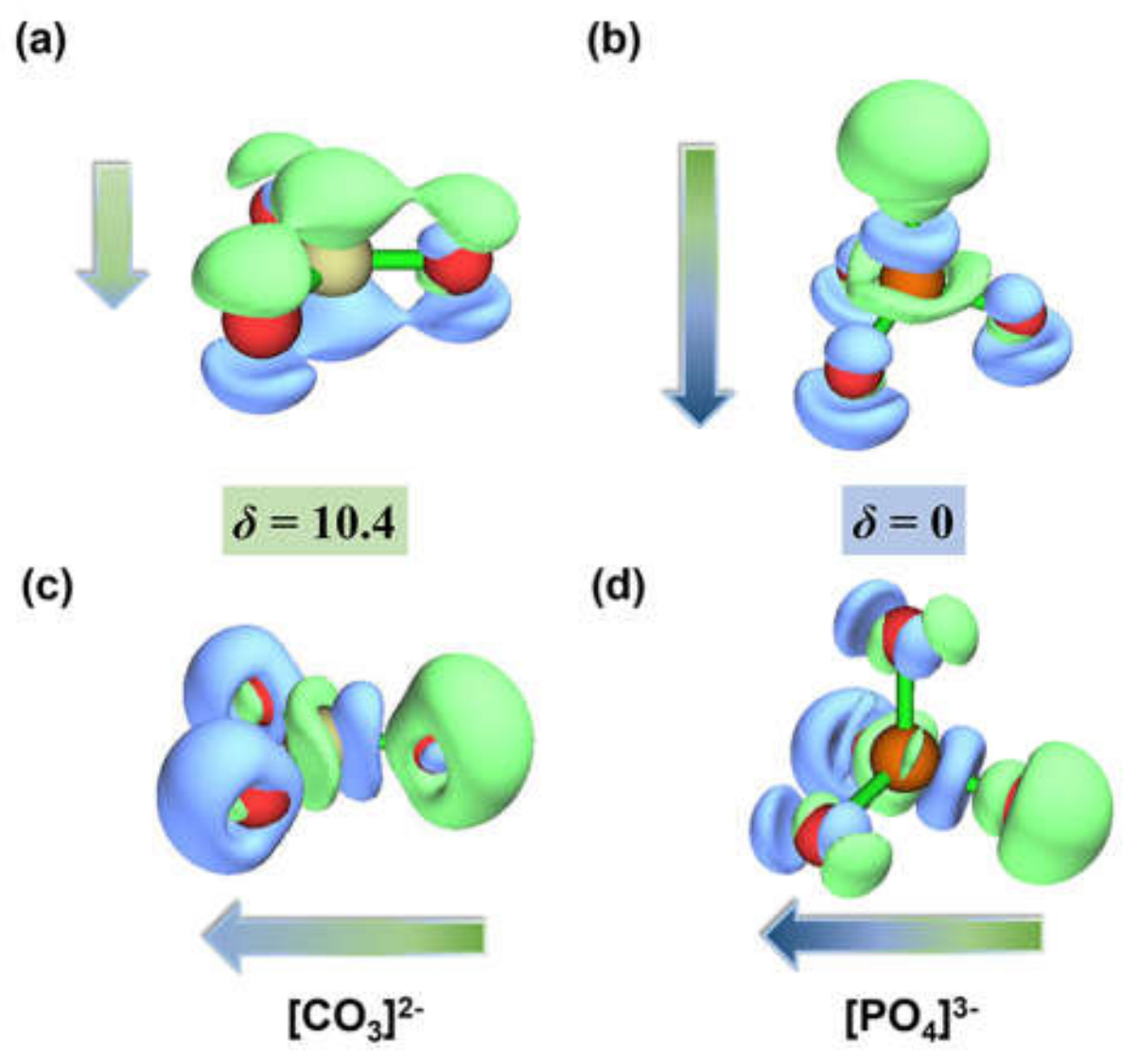

**Figure 2**. Electron density distribution and dipole moment $\mu$ (debye) of (a, c) $[CO_3]^{2-}$ and (b, d) $[PO_4]^{3-}$ under an external electric field intensity of 0.02 a.u. along the out-plane (a, b) and in-plane (c, d) directions. (1 a.u. = 51.421 V/Å)

For the non-$\pi$-conjugated class, we consider $[PO_4]^{3-}$, because $[SO_4]^{2-}$ tends to release $SO_2$ gas at high temperature.[10] Besides, phosphates show a great structural diversity and crystal growth feasibility.[11] Taking all these into account, we select $[CO_3]^{2-}$ and $[PO_4]^{3-}$ to implement our principle as discussed below. First, to describe vividly the different isotropy nature of $[CO_3]^{2-}$ and $[PO_4]^{3-}$, an electric field with a strength of 0.02 a.u. (1 a.u. = 51.42 V/Å)[12] is imposed on the molecule as shown in **Figure 2**. The $[CO_3]^{2-}$ group exhibits anisotropic nature, with the induced dipole moments ($\mu$) are -12.7 D or -2.2 D, when the external electric field is parallel or perpendicular to the $[CO_3]^{2-}$ plane, giving a quantity difference of $\Delta|\mu|$ = 10.5 D that is 5.5 times greater than that of $[PO_4]^{3-}$ ($\Delta|\mu|$ = 1.9 D). Consistently, the polarizability anisotropy ($\delta$) of $[CO_3]^{2-}$ overwhelms that of $[PO_4]^{3-}$ ($\delta$ = 10.4 vs 0.0). (**Figure 2**) Thus, the $[CO_3]^{2-}$ group is much more anisotropic than the $[PO_4]^{3-}$ group. Therefore, the optical anisotropic of carbonophosphate mainly comes from the $\pi$-conjugated $[CO_3]^{2-}$ group of which the $\pi$ electrons of $[CO_3]^{2-}$ are more flexible and polarizable along the in-plane direction that drag the phonon propagation leading to a much larger in-plane refractive index and a smaller out-plane one, and eventually generates a large birefringence (Δ*n*) for the carbonophosphate.

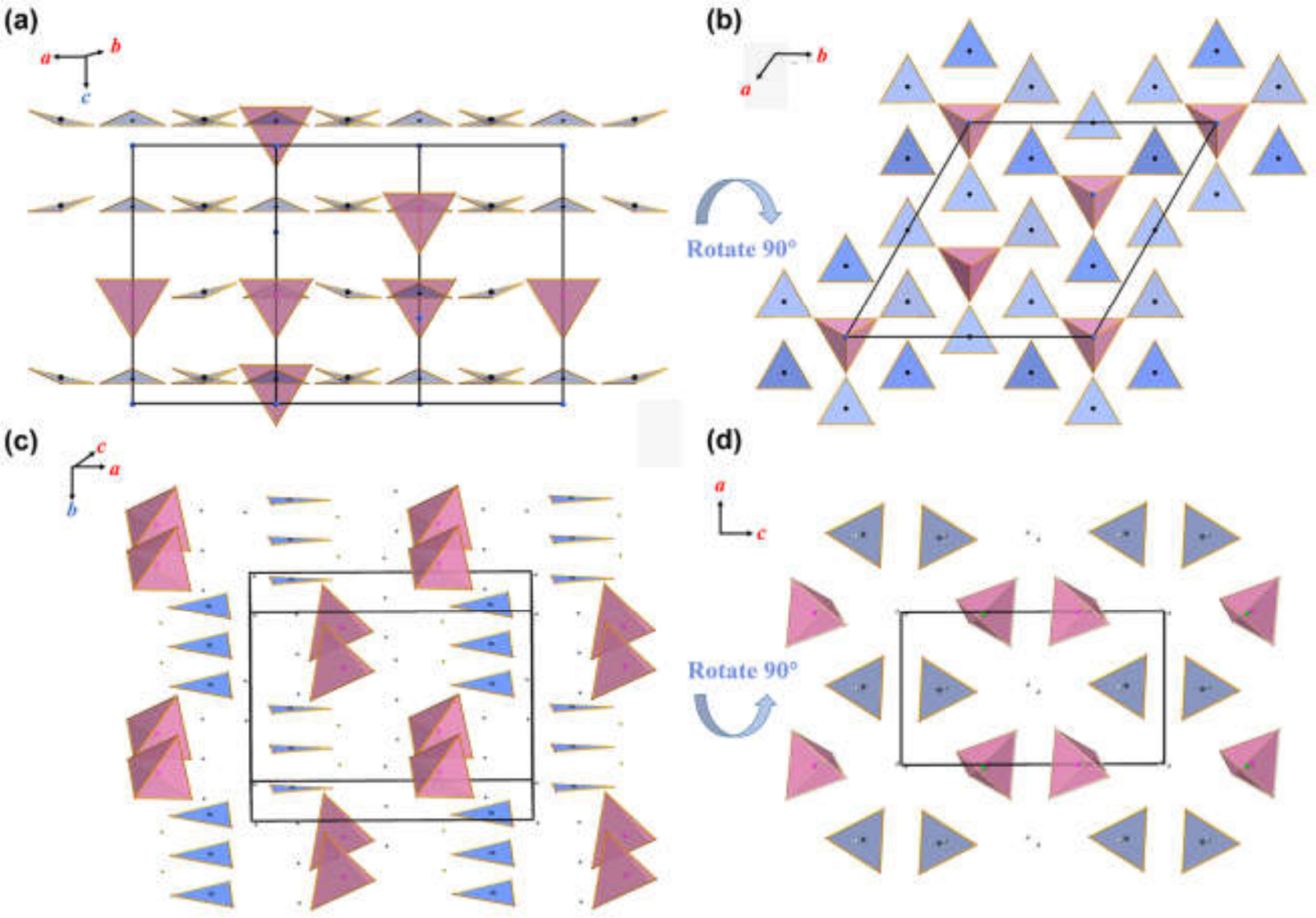

**Figure 3**. Crystal structures of (a, b) **SYPC₃** and (c, d) **NSrPC**. The polyhedron: blue, $[CO_3]^{2-}$ triangle; pink, $[PO_4]^{3-}$ tetrahedron.

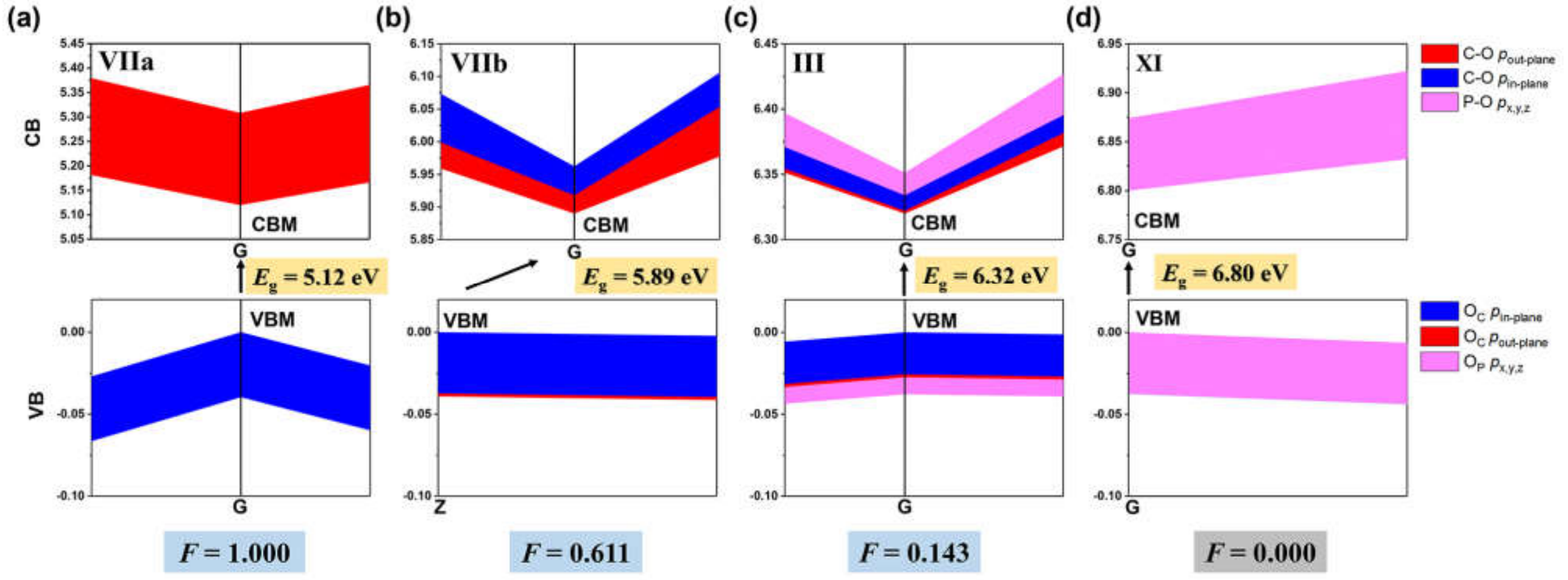

**Figure 4**. The orbital projected CB and VB at the minimum along the $k_z$ direction in the Brillouin zone of (a) $Na_2[CO_3]$ (**VIIa**) (b) $Na_2[CO_3]$ (**VIIb**), (c) **NSrPC** (**III**), (d) $NaSr[PO_4]$ (**XI**), respectively. The *p* states of the $[CO_3]^{2-}$ and $[PO_4]^{3-}$ are projected on the VBM and CBM, and the curve width represents the orbital volume density (1/Å³). Additional figures are presented in **Figures. S3, 4**

**Crystal structure feature and stability of carbonophosphate**

The mixture of $[CO_3]^{2-}$ and $[PO_4]^{3-}$, i.e., carbonophosphate, is a stable and experimentally feasible system, which includes some of the naturally occurred mineral species, such as Daqingshanite, found in China.[13] There are approximately 100 carbonophosphates that are identified crystallographically according to the inorganic crystal structure database, but only 5 of them crystallize as an NCS structure, none of them has been studied for their optical properties to date. (**Table S1**)

$Sr_3Ce[PO_4][CO_3]_3$, a mineral Daqingshanite, crystalizes in a trigonal *R*3*m* structure constructed by both $[PO_4]^{3-}$ and $[CO_3]^{2-}$ groups.[13] (**Figure S1**) However, the *f*-*f* transition of the component Ce element is deleterious for the transparency in the DUV region, we thus construct an isostructural $Sr_3Y[PO_4][CO_3]_3$ (**SYPC₃**) containing no *f* electron. (**Figure S1**) The structural views in **Figure 3** emphasize the layer-like array of the isolated $[CO_3]^{2-}$ groups showing a dihedral angle between adjacent ones ($\gamma$ = 35.80°). The isolated $[PO_4]^{3-}$ tetrahedron is embedded in such a layer.

In a different structure type, $Na_3Sr[PO_4][CO_3]$ (**NSrPC**), also known as a mineral crawfordite,[14] crystallizes a monoclinic $P2_1$ structure constructed by the double-layered (010) arrays of

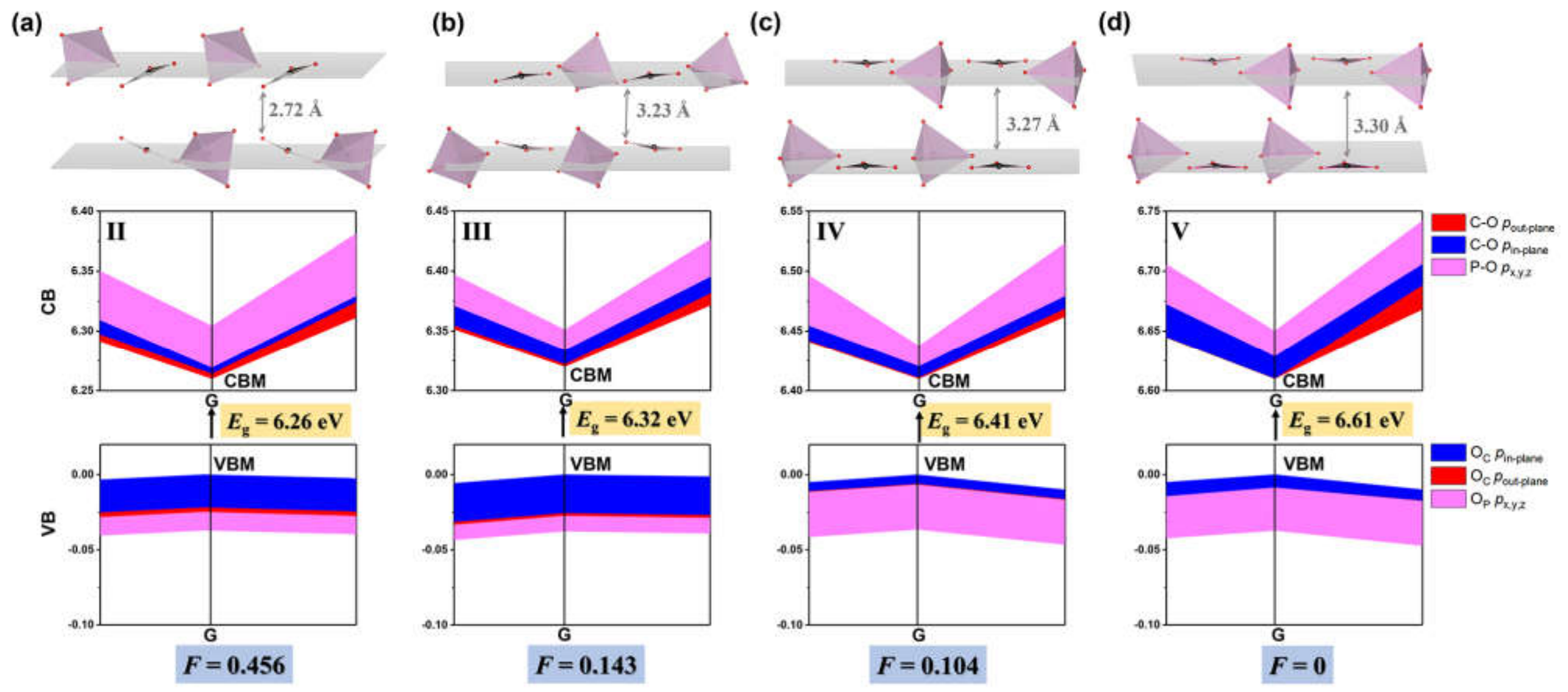

**Figure 5**. The inter-layer distance, orbital projected CB and VB at the minimum along the $k_z$ direction in the Brillouin zone of (a) **NBaPC** (**II**), (b) **NSrPC** (**III**) (c) **NCaPC** (**IV**), and (d) **NMgPC** (**V**), respectively. The *p* states of the $[CO_3]^{2-}$ and $[PO_4]^{3-}$ projected on the VBM and CBM, and the curve width represents the orbital volume density (1/Å$^3$).

individual $[CO_3]^{2-}$ groups. These arrays are separated by the isolated $[PO_4]^{3-}$ tetrahedra, alternatively. Note that the neighboring pair of $[CO_3]^{2-}$ groups are antiparallel aligned. (**Figure 3c, d**) Three more isostructural $Na_3X[PO_4][CO_3]$, (**NXPC**, X = Ba, Ca, Mg), are also built and studied. (**Table S2**)

**Identifying the enhancing mechanism of Δ*n* and $E_g$ of carbonophosphates**

Before we study the electronic structures and the linear and nonlinear optical properties of **SYPC₃** and **NXPC** (X = Ba, Sr, Ca, Mg), we first study their stability. The phonon vibrational spectra (**Figure S2**) reveal no imaginary phonon mode indicating both **SYPC₃** and **NXPC** are kinetically stable. In addition, the calculated elastic coefficients demonstrate their mechanical stability.[15] (**Table S3**) Secondly, we confirm the reliability and accuracy of our calculation by running the test calculations on the benchmark KDP and other 10 representative carbonates and phosphates whose experimental optical property data are available. With the aid of Vienna *Ab-initio* Simulation Package (VASP),[16] the $E_g$ values were obtained by the Heyd-Scuseria-Ernzerhof (HSE06) hybrid functional based on a screened Coulomb potential,[17] our calculations realize an excellent consistency with the experimental observations. (**Table S4**)

Subsequently, the optical properties of **SYPC₃** (**I**) and **NXPC** (**II–V**) together with the selected carbonates ($Na_3Y[CO_3]_3$ (**VI**), $Na_2[CO_3]$ (**VIIa**, **VIIb**); $Sr[CO_3]$ (**VIII**); $Na_2Mg[CO_3]_2$ (**IX**)) and phosphates ($Sr_3[PO_4]_2$ (**X**), $NaSr[PO_4]$ (**XI**), $NaMg[PO_4]$ (**XII**), $Sr_3Y[PO_4]_3$ (**XIII**), $Na_3Y[PO_4]_2$ (**XIV**)) are calculated. (**Table 1**)

As listed in **Table 1**, the phosphate showing weak anisotropy exhibits a small Δ*n* ranging from 0.005 to 0.016; whereas, the carbonophosphate exhibits a greatly enhanced optical anisotropy with Δ*n* that is 3–24 times higher falling in a range (0.048 to 0.121) that is comparable to those of the *π*-conjugated carbonates. For instance, **NSrPC** (**III**, Δ*n* = 0.088) vs $NaSr[PO_4]$ (**XI**, Δ*n* = 0.006). Simultaneously, the $E_g$ of carbonophosphate is widened by 0.2 to 1.7 eV in comparison with those of carbonates. Such as **SYPC₃** (**I**, $E_g$ = 6.85 eV) vs $Na_3Y[CO_3]_3$ (**VI**, $E_g$ = 5.59 eV); **NSrPC** (**III**, $E_g$ = 6.32 eV) vs $Na[CO_3]$ (**VIIa**, $E_g$ = 5.12 eV). Significantly, as the band gap of the carbonophosphate is wider, which ensures a shorter absorption cutoff edge that falls in the DUV region (181–196 nm), about 20–40 nm shorter than those of carbonates. According to the refractive index dispersion curve (**Figure 7a**), the $\lambda_{PM}$ of **SYPC₃** (**I**) (181 nm) is the shortest among phosphates, which is 78 nm shorter than that of KDP. It should note that except **NBaPC** (**II**, Δ*n* is only 0.048), compounds **I**, and **III–V** can achieve phase-matching in whole transparent range. More significantly, **SYPC₃** (**I**) exhibits a large static $d_{21}$ of 2.029 pm/V, about 3 times stronger than those of KDP ($d_{36(cal.)}$ = 0. 76 pm/V) and KBBF ($d_{11(cal.)}$ = 0. 61 pm/V). (**Table 1**) For **NXPC** (**II–V**), the $d_{ij}$ is moderately strong, ranging from 0.2–0.9 times that of KDP. In order to fully understand the micro mechanism, detailed analyses and discussions are presented as following.

**The *π*-conjugated confinement maximizes the band gap of carbonophosphate**

As shown in **Figure S3**, for $Sr[CO_3]$ (**VIII**), the valence band maximum (VBM) is dominated by the in-plane *p* bands, whereas the conduction band minimum (CBM) is dominated by the out-plane $p_\pi^*$ bands. On the contrary, in the non-*π*-conjugated $Sr_3[PO_4]_2$ (**X**), the *p* bands contribute in both VBM and CBM without any orientation preference. To quantitatively evaluate the $p_\pi^*$ contribution near the Fermi level, we define a factor of $F \stackrel{\text{def}}{=} D_{p_\pi}/D_{p_{total}}$ at the CBM, where *D* means the orbital volume density; for a non-*π* conjugated phosphate, the *F* is 0 indicating the *π*-conjugated coupling is no longer existing. As shown by the selected 15 compounds, the larger the *F* is, the narrower the $E_g$. (**Table 2**) Generally, when *F* = 1, the *π*-bond coupling contributes the most, under which the band gap is the narrowest. Note that the *F* value is firstly influenced by the coplanarity of the $[CO_3]^{2-}$ groups. For example, for the two $Na_2[CO_3]$ (**VIIa**, **VIIb**), as the perfect coplanarity disappears from **VIIa** to **VIIb**, the *F* value drops from 1.000 to 0.611, reflecting a less contribution of the $p_\pi^*$ states,

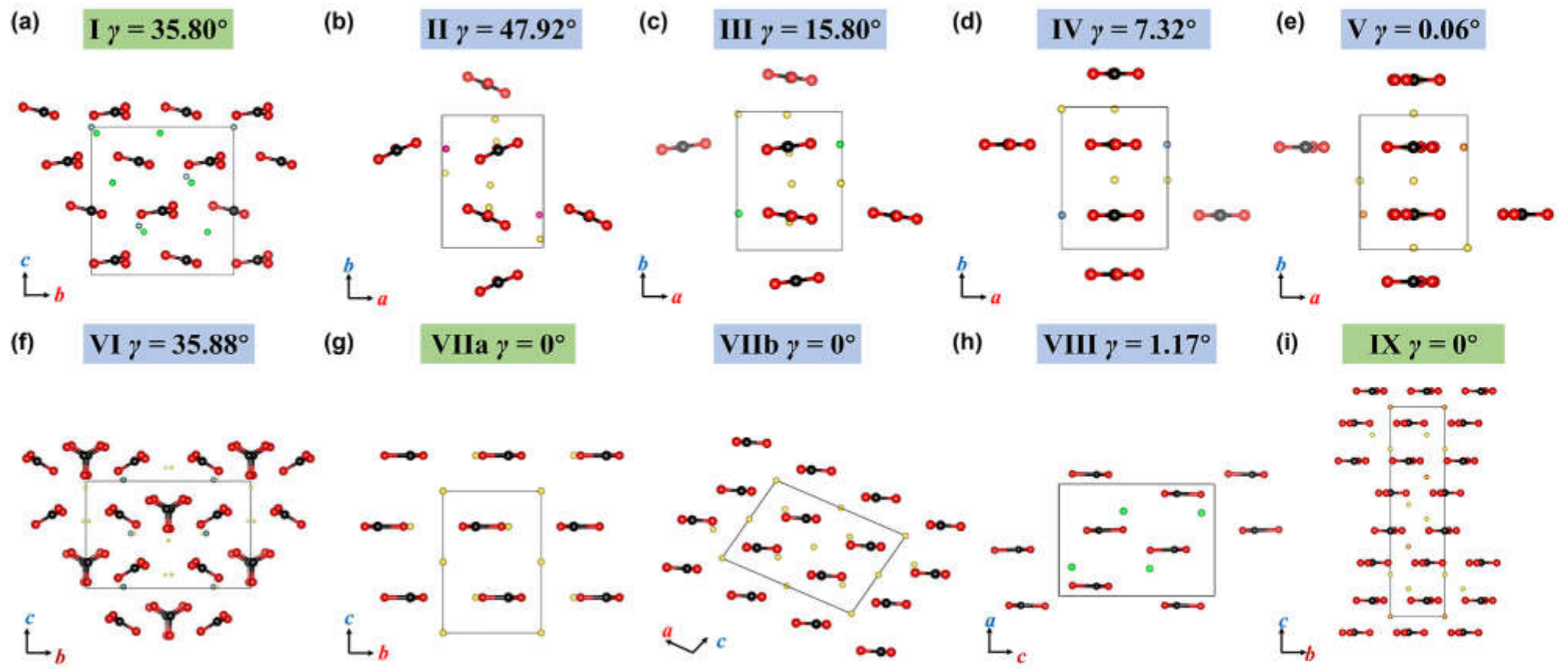

**Figure 6.** Simplified crystal structures (with the $[PO_4]^{3-}$ group omitted) (a) **SYPC₃** (**I**) (b-e) **NXPC** (X = Ba (**II**), Sr (**III**), Ca (**IV**), Mg (**V**)), (f) $Na_3Y[CO_3]_3$ (**VI**), (g) $Na_2[CO_3]$ (**VIIa**), $Na_2[CO_3]$ (**VIIb**), (h) $Sr[CO_3]$ (**VIII**), (i) $Na_2Mg[CO_3]_2$ (**IX**). The dihedral angle (γ) is marked. Green: uniaxial crystal, blue: biaxial crystal.

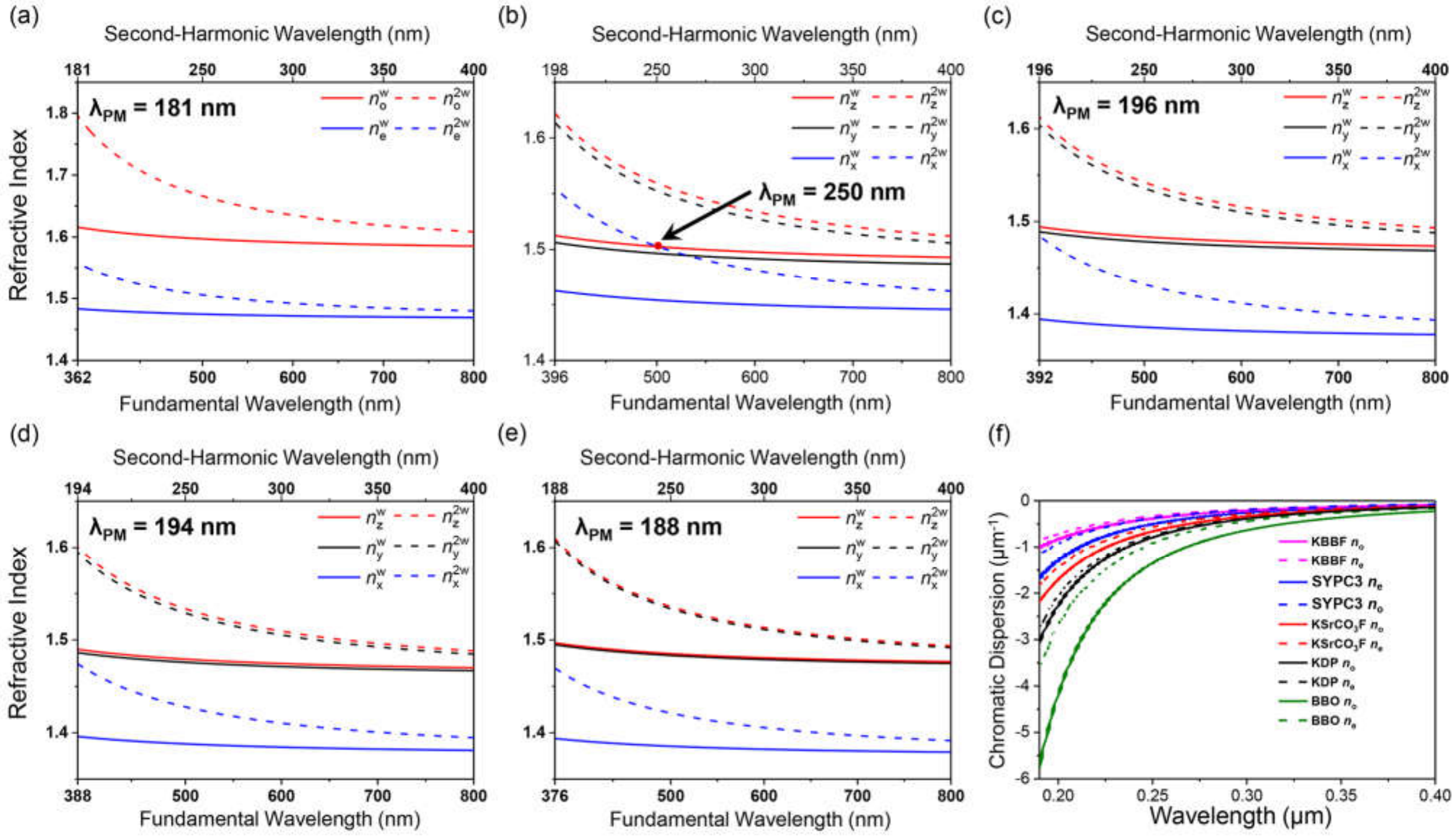

**Figure 7.** Calculated type I phase-matching condition of (a) **SYPC₃** (**I**) and (b-e) **NXPC** (X = Ba (**II**), Sr (**III**), Ca (**IV**), Mg (**V**)). The $\lambda_{PM}$ is determined when $n_e^{2w} = n_o^w$(or $n_x^{2w} = n_z^w$). Noted that when the shortest second-harmonic wavelength is calculated to be shorter than $\lambda_{cutoff}$, $\lambda_{PM}$ is set to be = $\lambda_{cutoff}$. The chromatic dispersions of **I** compared with those of KBBF, $KSrCO_3F$, KDP and BBO (f).

which is associated with an $E_g$ increase from 5.12 to 5.89 eV. (**Figure 4a**, **b**) Secondly, the *F* value is effectively reduced by the involvement of $[PO_4]^{3-}$ group. For instance, $Na_3Y[CO_3]_3$ (**VI**, 5.59 eV) vs **SYPC₃** (**I**, 6.85 eV), and $Na_2Mg[CO_3]_2$ (**IX**, 6.00 eV) vs **NMgPC** (**V**, 6.60 eV). As depicted in **Figure 4**, the VBM of **NSrPC** (**III**) looks like a simple mixture of $Na_2[CO_3]$ (**VIIb**) and $NaSr[PO_4]$ (**XI**), yet the CBM shows a sharply decline of the $p_\pi^*$ component, leading to a small *F* = 0.143, and consequently a large $E_g$ = 6.32 eV. Evidently, the separation of $[PO_4]^{3-}$ intercepts the delocalized *π-π* interactions of the $[CO_3]^{2-}$ groups, which results in an $E_g$ enhancement ($E_{g,\ carbonophosphate} > E_{g,\ carbonate}$). Consistently, **NXPC** (**II–V**) reveal that as the increasing of the interlayer distance from 2.72 to 3.30 Å, $E_g$ enhances from 6.26 to 6.61 eV. (**Figure 5**)

**The birefringence dependence on the density and coplanarity of $[CO_3]^{2-}$ group**

Judging from the crystal symmetry, the trigonal **SYPC₃** (**I**), $Na_2Mg[CO_3]_2$ (**IX**) and $Sr_3[PO_4]_2$ (**X**), together with the hexagonal $Na_2[CO_3]$ (**VIIa**), belong to the uniaxial crystal class, which exhibits two unequal principal refractive indices known as: the extraordinary ($n_e$) and ordinary ($n_o$) refractive indices, respectively. And the other eleven compounds, monoclinic **NXPC** (**II**–**V**) and $Na_2[CO_3]$ (**VIIb**), orthorhombic $Na_3Y[CO_3]_3$ (**VI**), $Sr[CO_3]$ (**VIII**), $NaSr[PO_4]$ (**XI**), $NaMgPO_4$ (**XII**) and $Na_3Y[PO_4]_2$ (**XIV**) belong to the biaxial crystal class, which exhibits three distinct refractive indices of $n_z > n_y > n_x$. (**Table 1**)

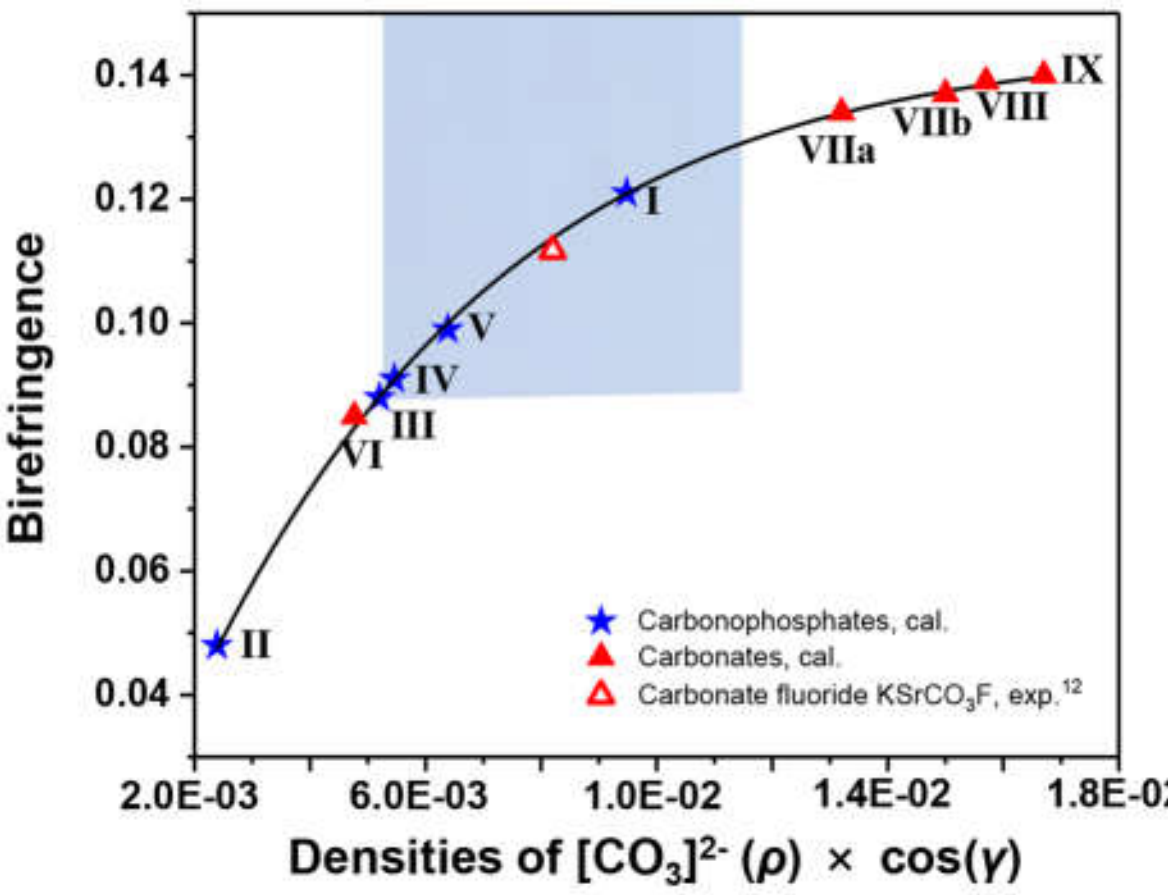

**Figure 8.** Boltzmann function of the birefringence-density of $[CO_3]^{2-}$-dihedral angle dependence.The compounds and detailed data are listed in **Table 2**.

Previous studies on the centrosymmetric and noncentrosymmetric solids, for example, in the simple carbonates, nitrates and borates, show that the refractivity depends deeply to the arrangement of the $\pi$-conjugated group.[18] And the birefringence is determined by the coplanarity and the density of the $\pi$-conjugated $[BO_3]^{3-}$ groups in simple borates such as $TM_3(BO_3)_2$ (TM = Zn, Cd)[19], $Cd_2B_2O_5$[20], and $M_3(BO_3)_2$ (M = Hg, Mg, Ca, Sr)[21]. Similar situation is found in carbonophosphate, in which the $[CO_3]^{2-}$ density per unit cell is denoted as $\rho$, and the coplanarity of $[CO_3]^{2-}$ is measured by the dihedral angle ($\gamma$) between two neighboring $[CO_3]^{2-}$ groups. For carbonate **VIIa**, **VIIb**, **VIII**, and **IX** (**Figure 6g**–**i**), despite of the different crystallographic symmetries, the $[CO_3]^{2-}$ groups are arranged in a coplanar manner ($\gamma$ = 0° or 1.17°) giving rise to a relatively large birefringence. Besides, as $\rho$ increases from 1.32 to 1.67 ($\times 10^{-2}$ $n$/Å$^3$), $\Delta n$ increases from 0.134 to 0.140. Similarly, as $\rho$ increases from **II** to **V**, $\Delta n$ increase from 0.048 to 0.099, for which the increase of the coplanarity ($\gamma$ decreases from 47.92 to 0.06°, **Figure 6b**–**e**) shall also contributes. (**Table 2**) Further, the statistical analyses reveal that the dependence of birefringence ($\Delta n$)-density of $[CO_3]^{2-}$ ($\rho$)-dihedral angle ($\gamma$) obeys a Boltzmann function (**equ. 1**) with an excellent goodness of fit (R2 = 0.999).

$$\Delta n = 0.14532 - \frac{1.0045\ 1}{1+e^{\frac{\rho*\cos\gamma+0.00837}{0.00484}}} \quad (1)$$

Such a $\Delta n$-$\rho$-$\gamma$ correlation points out that when seeking for large birefringence, one should first consider not only how to maintain the density of the $\pi$-conjugated moiety as high as possible; but also the coplanarity as perfect as possible. As shown in **Figure 8**, **NMgPC** (**V**) has a large $\Delta n$ of about 0.099 when the $[CO_3]^{2-}$ groups achieve nearly perfect coplanarity with $\gamma$ = 0.06°. To go beyond this limit, **SYPC₃** (**I**) reveals enhancing further the density of $[CO_3]^{2-}$ ($\rho$) is possible to increase $\Delta n$ to 0.121. Note that **Figure 6a** shows the $[CO_3]^{2-}$ groups in **I** are poorly coplanar with a $\gamma$ = 35.80°, therefore, one can infer that under the same density of the $[CO_3]^{2-}$ groups, there is still a large room to enhance $\Delta n$ via improving the coplanarity. The highlighted area in **Figure 8** points out the possible direction for the future exploration of DUV NLO carbonophosphates. In addition, **SYPC₃** displays a small chromatic dispersion that is favorable to achieve a short $\lambda_{PM}$. The dispersion index,[22] which is defined as the derivative of the refractive index to the wavelength ($dn/d\lambda$), of **SYPC₃** is smaller than those of the representative materials except for KBBF. (**Figure 7f**)

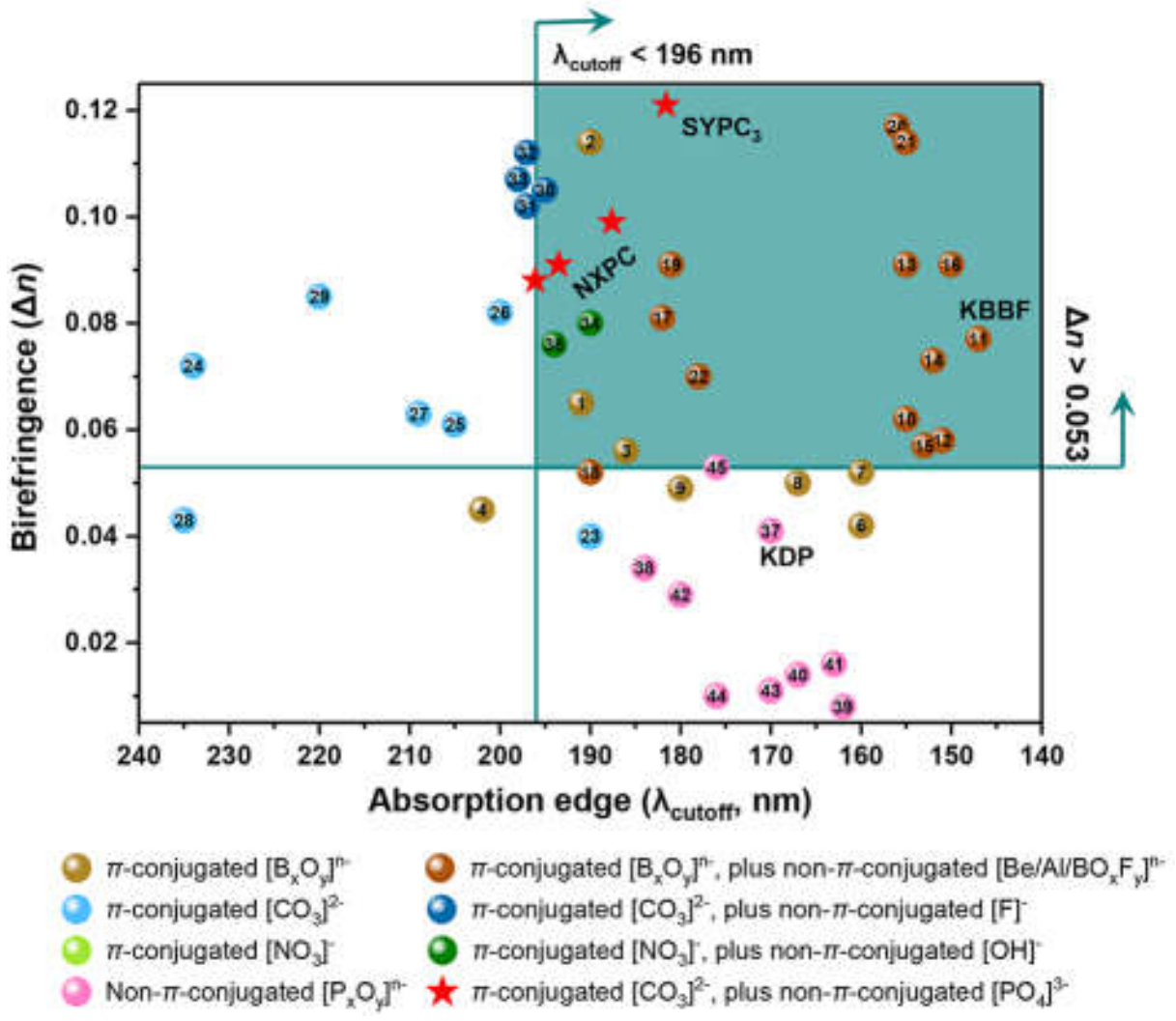

**Figure 9**. Distribution of representative materials with respect to the absorption edge ($\lambda_{cutoff}$) and birefringence ($\Delta n$). The highlighted first quadrant represents those qualified as a DUV NLO candidate showing $\lambda_{PM}$ < 200 nm and $\Delta n$ > 0.053, simultaneously. The 45 selected known compounds are detailed in **Table S5**.

**The generality of the $\pi$-conjugated confinement**

The 1st quadrant on **Figure 9** represents the preferred area ($\lambda_{PM}$ < 200 nm) for DUV NLO materials where the highly desired balance between $\Delta n$ and $\lambda_{cutoff}$ are achieved, which is nearly solely occupied by borates to date. Remarkably, all of these materials are in fact "$\pi$-conjugated confinement" structures. For instance, in the KBBF structure, the $\pi$-conjugated $[BO_3]^{2-}$ groups are in fact separated by the non-$\pi$-conjugated $[BeO_3F]^{5-}$ tetrahedra;[4] therefore, the $\pi$-conjugated interactions between $[BO_3]^{2-}$ groups are confined within the boundary that is defined by the separation of the non-$\pi$ $[BeO_3F]^{5-}$. Such a $\pi$-conjugated confinement leads to an $E_g$ enhancing as discussed above. More examples are $A_2NO_3[OH]_3$ (**34, 35**), where the $\pi$-conjugated $[NO_3]^-$ groups are separated by the non-$\pi$-conjugated hydroxide $[OH]^-$ group.[23] Without such a separation, the $E_g$ of the $Rb_2Na(NO_3)_3$ counterpart is too narrow, giving a long $\lambda_{cutoff}$ of 270 nm, which is beyond the range shown. Very interestingly, from $Na_3Y[CO_3]_3$ (**29**) to $ASrCO_3F$ (**30**–**33**), the separation of $[CO_3]^{2-}$ groups by the non-$\pi$-conjugated $[F]^-$ anion pushes the latter into the 1st quadrant, although marginally. Further, from $ASrCO_3F$ (**30**–**33**) to **SYPC₃**, with larger non-$\pi$-conjugated $[PO_4]^{3-}$ anions, the separation becomes more efficient and gives rise to a more profound $\pi$-conjugated confinement which eventually results in a further significant property improvement on both $E_g$ and $\Delta n$. (**Table S5**) Our theory provides a clearer direction, better experimentally feasibility and manipulability to pave the way towards the design and searching of new DUV NLO materials that otherwise relies on try-and-error experience. And much energy and money will be saved.

## Theoretical Methodology

**_Ab-initio_ calculations of the _π_-conjugated $[TO_3]^{x-}$ and non-_π_-conjugated $[MO_4]^{x-}$ groups**

The Gaussian 09W program package[24] was employed to perform the calculations on structure geometry optimization and HOMO–LUMO gap estimation by the Becke's 3-parameter hybrid exchange functional (B3LYP) with 6-31G(d) basis set.[25] The polarizability anisotropy ($\delta$) and first hyperpolarizability ($|\beta_{max}|$) were calculated by the CAM-B3LYP hybrid functional with aug-cc-pVDZ basis set.[26] The external electric field is set to be 0.02 a.u. (1.03 V/Å) for guaranteeing self-consistent field iteration converged. [12]

**First-Principles calculations of the $Sr_3Y[PO_4][CO_3]_3$ and $Na_3X[PO_4][CO_3]$ crystals**

The Vienna *Ab-initio* Simulation Package (VASP) was used to calculate the electronic structure as well as the linear and nonlinear optical properties.[16] The generalized gradient approximation (GGA) in the scheme of Perdew-Burke-Eruzerhof (PBE) was used to describe the electron-ion core interactions with a plane-wave energy cutoff of 600 eV.[27] A Monkhorst-pack *k*-point meshes spanning less than 0.02/$Å^3$ in the Brillouin zone was used for the crystal geometry optimization and electronic structure calculations.[28]

**Optical properties calculations and accurate band gap estimation**

Based on the optimized geometry structure, the imaginary part of the dielectric function was calculated, and the real part of the dielectric function was determined by using the Kramers-Kronig transform,[29] and then the refractive index ($n$) and birefringence ($\Delta n$) were obtained. The shortest SHG output wavelength ($\lambda_{PM}$) was calculated based on the dispersion curves of the refractive index (e.g., $n_o$ and $n_e$ for a negative uniaxial crystal) when $n_e$ ($2\omega$) = $n_o$ ($\omega$). When $\lambda_{PM}$ was shorter than $\lambda_{cutoff}$, $\lambda_{PM}$ was set equal to $\lambda_{cutoff}$. The so-called length-gauge formalism derived by Aversa and Sipe was adopted to calculate the NLO properties.[30] The static second-order nonlinear susceptibilities $\chi^{(abc)}$ ($-2\omega$, $\omega$, $\omega$) were calculated under the restriction of Kleinman's symmetry.[31] It should be emphasized that the GGA method with PBE functional usually underestimate the band gap. To calculate the optical properties accurately, a scissors operator was employed.[32] The band gap calculated by Heyd-Scuseria-Ernzerhof (HSE06) hybrid functional based on a screened Coulomb potential was used as the experimental band gap.[17] And the scissors operator was set as the difference between the HSE06 and GGA band gap. Subsequently, the scissors-corrected GGA method was employed to calculate the optical properties. Meanwhile, the linear response method was employed to calculate the phonon dispersion,[33] in which the LO-TO phonon frequency splitting at the Γ-point was also included. The dispersion separation of 0.01/$Å^3$ was adopted to make sure the good convergence. (for more computational method details, see the Supporting Information).

## Conclusion

In summary, we propose a general principle for the rational design of DUV NLO material, the *π*-conjugated confinement, to achieve the desired balance between large band gap ($E_g$) and large birefringence ($\Delta n$), which are otherwise irrelated and even reversely correlated. The π-conjugated confinement not only widens $E_g$ by the separation of the non-π-conjugated groups, but also keeps a suitable density of the *π*-conjugated groups so as to maintain a large optical anisotropy. Such a principle is valid for all of the known DUV NLO materials with favorable properties to date, and thus, it provides an essential design criterion. (**Figure 9**) Guided by this principle, we predict carbonophosphate as an excellent DUV NLO candidate. Two series of carbonophosphates, $Sr_3Y[PO_4][CO_3]_3$ (**SYPC$_3$**, **I**), and $Na_3X[PO_4][CO_3]$ (**NXPC**, **II**–**V**), exhibit not only improved $\Delta n$ values that are 3–24 times larger than that of simple phosphate, but also enlarged $E_g$ values that are 0.2–1.7 eV wider than that of simple carbonate. Significantly, **SYPC$_3$** distinguishes itself with excellent properties with a $\Delta n$ = 0.121, SHG laser output with $\lambda_{PM}$ = 181 nm that is 78 nm shorter than that of KDP, being the shortest among phosphates to date. Instructively, the Boltzmann function presented in **equ. 1** explains and predicts the correlation between the birefringence and the density and coplanarity of the *π*-conjugated $[CO_3]^{2-}$ group. The future experimental works on this and other possible combinations of all the conventional *π*-conjugated $[TO_3]^{x-}$ (T = B, C, N) and non-*π*-conjugated $[MO_4]^{x-}$ building units are in great anticipation.

## Acknowledgements

This research was supported by the National Natural Science Foundation of China under Project (21971019), and by Beijing Natural Science Foundation (2202022).

**Conflict of interest**

The authors declare no conflict of interest.

**Keywords:** *π*-conjugated • nonlinear optical materials • birefringence • the first principle calculation • phase-matching

**Table 1:** Calculated and experimental linear and nonlinear optical properties of selected compounds with commercial $KH_2[PO_4]$ (KDP) as a reference.

| | Carbonophosphate | | | | | Reference |
|---|---|---|---|---|---|---|
| **Compound** | $Sr_3Y[PO_4][CO_3]_3$ **($SYPC_3$, I)** | $Na_3Ba[PO_4][CO_3]$ **(NBaPC, II)** | $Na_3Sr[PO_4][CO_3]$ **(NSrPC, III)** | $Na_3Ca[PO_4][CO_3]$ **(NCaPC, IV)** | $Na_3Mg[PO_4][CO_3]$ **(NMgPC, V)** | $KH_2PO_4$ $K[PO_2OH_2]$, KDP |
| **Space group, linear optical classification** | *R*3*m*, uniaxial | $P2_1$, biaxial | | | | $I\bar{4}2d$, uniaxial |
| **$E_g$[a] ($\lambda_{cutoff}$, nm)** | 6.85 (181.0) | 6.21 (198.1) | 6.32 (196.2) | 6.41 (193.4) | 6.61 (187.6) | 7.03 (196.4) |
| **$\Delta n$[b]** | 0.121 | 0.048 | 0.088 | 0.091 | 0.099 | 0.040 |
| **Static SHG tensors (pm/V)** | $d_{15}$ = 0.508, $d_{21}$ = 2.029, $d_{33}$ = 0.363 | $d_{21}$ = -0.248 $d_{22}$ = -0.710 $d_{23}$ = 0.571 | $d_{21}$ = -0.209 $d_{22}$ = -0.387 $d_{23}$ = 0.320 | $d_{21}$ = -0.148 $d_{22}$ = -0.228 $d_{23}$ = 0.153 | $d_{21}$ = -0.102 $d_{22}$ = -0.157 $d_{23}$ = 0.114 | $d_{36}$ = 0.760 |
| **$I^{cal}/I^{cal}_{KDP}$[c]** | 2.7 × KDP | 0.9 × KDP | 0.5 × KDP | 0.3 × KDP | ~ 0.2 × KDP | |
| **$\lambda_{PM}$ (nm)** | 181.0 | 249.9 | 196.2 | 193.5 | 187.6 | 259.0 |

| | Carbonate | | | | | Phosphate | | | | |
|---|---|---|---|---|---|---|---|---|---|---|
| **Compound** | $Na_3Y[CO_3]_3$ **(VI)** | $Na_2[CO_3]$ **(VIIa, VIIb)** | | $Sr[CO_3]$ **(VIII)** | $Na_2Mg[CO_3]_2$ **(IV)** | $Sr_3[PO_4]_2$ **(X)** | $NaSrPO_4$ **(XI)** | $NaMgPO_4$ **(XII)** | $Sr_3Y[PO_4]_3$ **(XIII)** | $Na_3Y[PO_4]_2$ **(XIV)** |
| **Space group, linear optical classification** | *Ama*2, biaxial | $P6_3/mmc$, uniaxial | *C*2/*m*, biaxial | *Pnma*, biaxial | $R\bar{3}$, uniaxial | $R\bar{3}m$, uniaxial | *Pnma*, biaxial | $P2_12_12_1$, biaxial | $I\bar{4}3d$, isotropical | $Pca2_1$, biaxial |
| **$E_g$[a]** | 5.59 | 5.12 | 5.89 | 5.85 | 6.00 | 7.53 | 6.80 | 7.03 | 7.28 | 7.03 |
| **$\Delta n$[b]** | 0.085 | 0.134 | 0.137 | 0.139 | 0.140 | 0.016 | 0.006 | 0.005 | / | 0.013 |

[a] calculated by HSE06 hybrid functional. [b],[c] calculated at 532 nm.

**Table 2.** Birefringence, density of $[CO_3]^{2-}$, dihedral angle between two $[CO_3]^{2-}$ groups, *F* factor and $E_g$ for selected carbonophosphate, carbonate and phosphate.

| Compound | | $\Delta n$-related | | | $E_g$-related | | |
|---|---|---|---|---|---|---|---|
| | | **$\rho$ (×$10^{-2}$ $n$/Å$^3$)[a]** | **$\gamma$ (°)[b]** | **$\Delta n$[c]** | **$F$[d]** | **$d$ (Å)[e]** | **$E_g$ (eV)[f]** |
| Carbonophosphate[g] | $Sr_3Y[PO_4][CO_3]_3$ (**I**) | 1.174 | 35.80 | 0.121 | 0.170 | 2.978 | 6.85 |
| | $Na_3Ba[PO_4][CO_3]$ (**II**) | 0.531 | 47.92 | 0.048 | 0.456 | 2.722 | 6.26 |
| | $Na_3Sr[PO_4][CO_3]$ (**III**) | 0.540 | 15.80 | 0.088 | 0.143 | 3.234 | 6.32 |
| | $Na_3Ca[PO_4][CO_3]$ (**IV**) | 0.583 | 7.32 | 0.091 | 0.104 | 3.271 | 6.41 |
| | $Na_3Mg[PO_4][CO_3]$ (**V**) | 0.638 | 0.06 | 0.099 | 0 | 3.304 | 6.61 |
| Carbonate[h] | $Na_3Y[CO_3]_3$ (**VI**) | 0.504 | 35.88 | 0.085 | 1.000* | 2.482 | 5.59 |
| | $Na_2[CO_3]$ (**VIIa**) | 1.322 | 0 | 0.134 | 1.000 | 3.250 | 5.12 |
| | $Na_2[CO_3]$ (**VII*b***) | 1.501 | 0 | 0.137 | 0.611 | 2.914 | 5.89 |
| | $Sr[CO_3]$ (**VIII**) | 1.569 | 1.17 | 0.139 | 0.996 | 2.995 | 5.85 |
| | $Na_2Mg[CO_3]_2$ (**IX**) | 1.669 | 0 | 0.140 | 0.971** | 3.024 | 6.00 |
| Phosphate | $Sr_3[PO_4]_2$ (**X**) | 0 | / | 0.016 | 0 | / | 7.60 |
| | $NaSr[PO_4]$ (**XI**) | | | 0.006 | | | 6.80 |
| | $NaMgPO_4$ (**XII**) | | | 0.005 | | | 7.03 |
| | $Sr_3Y[PO_4]_3$ (**XIII**) | | | / | | | 7.28 |
| | $Na_3Y[PO_4]_2$ (**XIV**) | | | 0.013 | | | 7.03 |

[a] density of $[CO_3]^{2-}$ group per unit cell ($n$/Å$^3$).

[b] dihedral angle ($\gamma$) measured between two $[CO_3]^{2-}$ groups.

[c] calculated at 532 nm.

[d] $F \stackrel{\text{def}}{=} Dp_{\text{out-plane}}/Dp_{\text{Total}}$ at the CBM, where *D* means the orbital volume density.

[e] distance between the two adjacent $[CO_3]^{2-}$ layers.

[f] calculated by HSE06 hybrid functional.

[g] **Figure 7** displays the refractive index dispersion curve.

[h] **Figure S5** shows the refractive index dispersion curve.

* as shown in **Figure S4(a)**, in **VI** , there has two types of $[CO_3]^{2-}$ groups that are almost orthogonal. The ones perpendicular to [100] showing perfect coplanarity on the *bc* plane, giving a $F = 1$, which shall represent the lowest energy gap.

** as shown in **Figure S4(d)**, in **VIII**, the $[CO_3]^{2-}$ is not strict coplanar that are significant distorted with a C atom shifting 0.02 Å from the O3 triangle plane, giving rise to a non-*π*-conjugated feature, consequently, despite a large *F* (0.971), the band gap is still large.

## For Table of Contents Only

## A General Principle: π-Conjugated Confinement Maximizes Band Gap of DUV NLO Materials

Lin Xiong, Li-Ming Wu,* and Ling Chen*

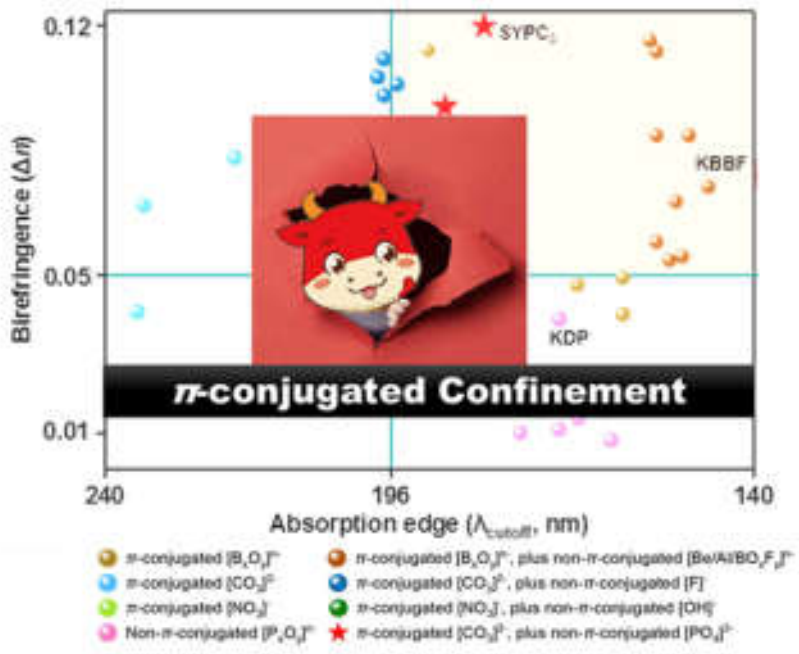

We propose a general principle, a π-conjugated confinement, for the rational design of DUV NLO materials whose band gaps are further enhanced in comparison with those of the simple π-conjugated salts. We reveal that the π-conjugated confinement is a shared structural feature for all the known DUV NLO materials (45 compounds) with favorable properties, and thus provides an essential design criterion. Guided by this, the carbonophosphate is discovered for the first time as a new promising DUV NLO candidate source. And **SYPC$_3$** exhibits the shortest SHG output laser among phosphates to date.

# Electronic Supplementary Information

# A General Principle: $\pi$-Conjugated Confinement Maximizes Band Gap of DUV NLO Materials

Lin Xiong, Li-Ming Wu,* and Ling Chen*

Beijing Key Laboratory of Energy Conversion and Storage Materials, Ministry of Education, College of Chemistry, Beijing Normal University, Beijing 100875 (People Republic of China)

**Abstract:** Current nonlinear optical materials face a conventional limitation on the tradeoff between band gap and birefringence, especially in the deep UV spectral region. To circumvent such a dilemma, we propose a general principle, a $\pi$-conjugated confinement, to partially decouple the inter group $\pi$-conjugated interactions with the separation of a non-$\pi$-conjugated group so as to maximize the band gap in comparison with those of simple $\pi$-conjugated salts, such as borates, carbonates. Meanwhile, to maintain a large optical anisotropy. We uncover that the $\pi$-conjugated confinement is a shared structural feature for all the known DUV NLO materials with favorable properties (45 compounds), and thus, it provides an essential design criterion. Guided by this principle, the carbonophosphate is predicted theoretically for the first time as a promising DUV candidate system, $Sr_3Y[PO_4][CO_3]_3$ and $Na_3X[PO_4][CO_3]$ (X = Ba, Sr, Ca, Mg) exhibit an enhanced birefringence that is 3–24 times larger than that of the simple phosphate, as well as an increased band gap that is 0.2–1.7 eV wider than that of the simple carbonate. Especially, the shortest SHG output of $Sr_3Y[PO_4][CO_3]_3$ is at $\lambda_{PM}$ = 181 nm, being the shortest one among phosphates to date.

**Table of Contents**

## Computation Method

To explore the geometry and electronic structure of the conventional $\pi$-conjugated planar $[TO_3]^{x-}$ (T = B, C, N) and non-$\pi$-conjugated tetrahedral $[MO_4]^{x-}$ (M = B, P, S) groups, systematic calculations were implemented via the Gaussian 09 Package with the hybrid B3LYP functional 6-31G(d) level. The dipole moment ($\mu$), polarizability anisotropy ($\delta$) and first hyperpolarizability ($|\beta_{max}|$) was calculated by CAM-B3LYP hybrid functional with aug-cc-pVDZ basis set. All wavefunction analyses and molecular structure maps were plotted by Multiwfn 3.7 code.[1] The following equation can be used to express a molecular system total energy which is under the effect of a homogeneous electric field (F) [2]

$$E = E^0 - \mu_i F_i - \frac{1}{2}\alpha_{ij}F_iF_j - \frac{1}{6}\beta_{ijk}F_iF_jF_k - \frac{1}{24}\gamma_{ijkl}F_iF_jF_kF_l - \cdots$$

In this equation, $E^0$ is the molecule energy without electronic field, $\mu_i$ refers to the dipole moment vector components, *F*, is the electric field, $\alpha_{ij}$ represents the tensor of the linear polarizability, the secondary and tertiary polarizability tensors are indicated as $\beta_{ijk}$ and $\gamma_{ijkl}$, respectively. Multiple subscripts *i*, *j*, and *k* represent *x*, *y*, and *z* coordinates. The derivate can be used to calculate the polarizability and hyperpolarizability analytically or numerically. The analytical third derivatives were employed to compute the static first hyperpolarizability $\beta_{max}$ because of more efficiency than numerical ones. Therefore, the following equations were used to calculate the polarizability anisotropy ($\delta$) and static first hyperpolarizability $|\beta_{max}|$ values: [3]

$$\delta = \sqrt{[(\alpha_{xx} - \alpha_{yy})^2 + (\alpha_{xx} - \alpha_{zz})^2 + (\alpha_{yy} - \alpha_{zz})^2]/2}$$

$$\beta_{max} = |\max|\beta_{ijk}\ i,j,k = \{x,y,z\}$$

The electronic structures and optical properties of $Sr_3Y[PO_4][CO_3]_3$ (**SYPC₃**) and $Na_3X[PO_4][CO_3]$ (**NXPC**, X = Ba, Sr, Ca, Mg) were calculated with the aid of the density functional theory (DFT) using the VASP code. The interactions between electrons and ion cores were represented by the norm-conserving pseudopotentials, and the valence electrons were treated as C-2$s^2$2$p^2$, O-2$s^2$2$p^4$, Na-2$p^6$3$s^1$, Mg-2$p^6$3$s^2$, P-3$s^2$3$p^3$, Ca-3$p^6$4$s^2$, Sr-4$p^6$3$d^{10}$5$s^2$, Y-4$p^6$4$d^1$5$s^2$, and Ba-5$p^6$4$d^{10}$6$s^2$. The generalized gradient approximation (GGA) in the scheme of Perdew-Burke-Eruzerhof (PBE) was used to describe the exchange and correlative potentials of the electron-ion interactions. Firstly, the atomic positions and lattice parameters were geometrically optimized with the aid of the Broyden-Fletcher-Goldfarb-Shannon (BFGS) algorithm.[4] An energy cutoff 600 eV and a *k*-point spacing with 0.05 $Å^{-1}$ were used. The atoms were allowed to relax until the forces on atoms were less than 0.01 eV $Å^{-1}$. Secondly, during the static self-consistent-field calculation, the plane-wave cutoff energy of 600 eV, threshold of $10^{-5}$ eV and dense Monkhorst–Pack *k*-point mesh spanning less than 0.02 $Å^{-1}$ performed by the tetrahedron method were employed.

Furthermore, it should be emphasized that the GGA method with PBE functional usually heavily underestimates the band gap $E_g$, whereas the Heyd-Scuseria-Ernzerhof (HSE06) hybrid functional based on a screened Coulomb potential was able to make an accurate prediction for the UV/DUV materials. Herein, the scissors-corrected GGA method was employed to calculate the optical properties. This self-consistent *ab initio* approach had been proved to be an efficient way for the investigation of the linear and nonlinear optical properties in many materials without introducing any experimental parameter. For the optical property calculation, additional unoccupied bands were added as the total number of electrons in the unit cell. Scissors operators of 1.94, 2.02, 1.67, 2.02, 2.09 and 1.62 eV for the band gap correction were applied for **$SYPC_3$**, **NXPC** (X = Ba, Sr, Ca, Mg) and $KH_2PO_4$ (KDP), respectively. According to Lin and Chen, the scissor value was formulated as

**Scissor value = band gap calculated by HSE06 – band gap calculated by GGA**

The dielectric function was defined as $\varepsilon(\omega) = \varepsilon_1(\omega) + i\varepsilon_2(\omega)$. Twice of the total number of valence bands were set, which was proved by the convergence testing to be sufficient for optical property calculations. The real part $\varepsilon_1(\omega)$ and the imaginary part $\varepsilon_2(\omega)$ were obtained from the momentum matrix elements between the occupied and unoccupied wave functions. The linear optical properties, such as refractive index $n(\omega)$, adsorption coefficient $\alpha(\omega)$ and reflectance coefficient $R(\omega)$, were obtained by the Kramers-Kroning transform. The refractive-index dispersive curves were configured by the least-squares method and the birefringences at variable wavelengths were acquired. The so-called length-gauge formalism derived by Aversa and Sipe was adopted to calculate NLO properties. At a zero frequency, the static second-order nonlinear susceptibilities can be described to purely interband ($\chi_e^{abc}$) and interband and intraband ($\chi_i^{abc}(-2\omega,\omega,\omega)$) processes.

$$\chi^{abc}(-2\omega,\omega,\omega) = \chi_e^{abc}(-2\omega,\omega,\omega) + \chi_i^{abc}(-2\omega,\omega,\omega)$$

where $\chi_e^{abc}(-2\omega,\omega,\omega)$ and $\chi_i^{abc}(-2\omega,\omega,\omega)$ are computed with the formulas as follows:

$$\chi_e^{abc} = \frac{1}{V}\sum_{nml,k} \frac{r_{nm}^a\{r_{ml}^b r_{\ln}^c\}}{\omega_{nm}\omega_{ml}\omega_{\ln}}[\omega_n f_{ml} + \omega_m f_{\ln} + \omega_l f_{nm}]$$

$$\chi_i^{abc} = \frac{i}{4V}\sum_{nm,k} \frac{f_{nm}}{\omega_{mn}^2}\left[r_{nm}^a\left(r_{mn;c}^b + r_{mn;b}^c\right) + r_{nm}^b\left(r_{mn;c}^a + r_{mn;a}^c\right) + r_{nm}^c\left(r_{mn;b}^a + r_{mn;a}^b\right)\right]$$

Here, superscripts *a*, *b*, and *c* are Cartesian components, $V = \frac{e^3}{h^2\Omega}$ is the unit cell volume, *r* is the position operator, $\hbar\omega_{nm} = \hbar\omega_n - \hbar\omega_m$, is the energy difference for the bands *m* and *n*, $f_{nm} = f_n - f_m$ is the difference of the Fermi distribution functions, $r_{nm;a}^b$ is the so-called generalized derivative of the coordinate operator in *k*-space,

$$r_{mn;a}^b = \frac{r_{nm}^a\Delta_{mn}^b + r_{nm}^b\Delta_{mn}^a}{\omega_{nm}} + \frac{i}{\omega_{nm}}\sum_l(\omega_{lm}r_{nl}^a r_{lm}^b - \omega_{nl}r_{nl}^b r_{lm}^a)$$

Where $\Delta_{mn}^a = (p_{nn}{}^a - p_{mm}{}^a)/m$ is the difference between the electronic velocities at the bands *n* and *m*. From these forms, it is obvious that both the $\chi_e^{abc}$ and $\chi_i^{abc}$separately automatically satisfy the full permutation, or Kleinman's symmetry.

## The Origin of SHG

The macroscopic NLO response was a collaborative effect of the microscopic NLO active units. **NSrPC** and **SYPC$_3$** were selected as examples to thoroughly investigate. The $d_{ij}$ contributions were thoroughly analyzed by the cutoff-energy-dependent SHG coefficient and the total and partial densities of states (DOS and PDOS), which were beneficial to understand the relationship between NCS structure and NLO activity. (**Figure S6**) The VB-1 of **NSrPC** ($E_F$ to -3.0 eV, 0.342 pm/V, 88.4%) consisted O 2*p* nonbonding states of $[CO_3]^{2-}$ and $[PO_4]^{3-}$ units, and CB-1 (6.3 to 7.3 eV, 0.236 pm/V, 61.0%) was dominated by the C-O hybrid $sp^2$ orbitals that made the most significant contributions to the SHG response ($d_{23}$). Their corresponding partial DOSs (PDOSs) and associated partial charge density maps in VB-1 and CB-1 showed the salient features as displayed in **Figure S6**c. Consequently, the C-O and P-O interactions were distributed in these key energy sections and thus contributed significantly to the $d_{23}$.

In case of **SYPC$_3$**, the states at VB-1 ($E_F$ to -3.8 eV, 1.863 pm/V, 92%), CB-1 (6.9 to 7.7 eV, 1.201 pm/V, 55%) and CB-2(7.3 to 10.0 eV, 1.166 pm/V, 36%) made the most significant contributions to the $d_{26}$. (**Figure S6**d, e) Additionally, the charge density map showed VB-1 and CB-1 were similar to those of **NSrPC**, while the CB-2 was mainly composed of Y 4*d* orbitals. Consequently the $[CO_3]^{2-}$, $[PO_4]^{3-}$, $[YO_{10}]$ units in **SYPC$_3$** all contributed to the SHG.

## Figures

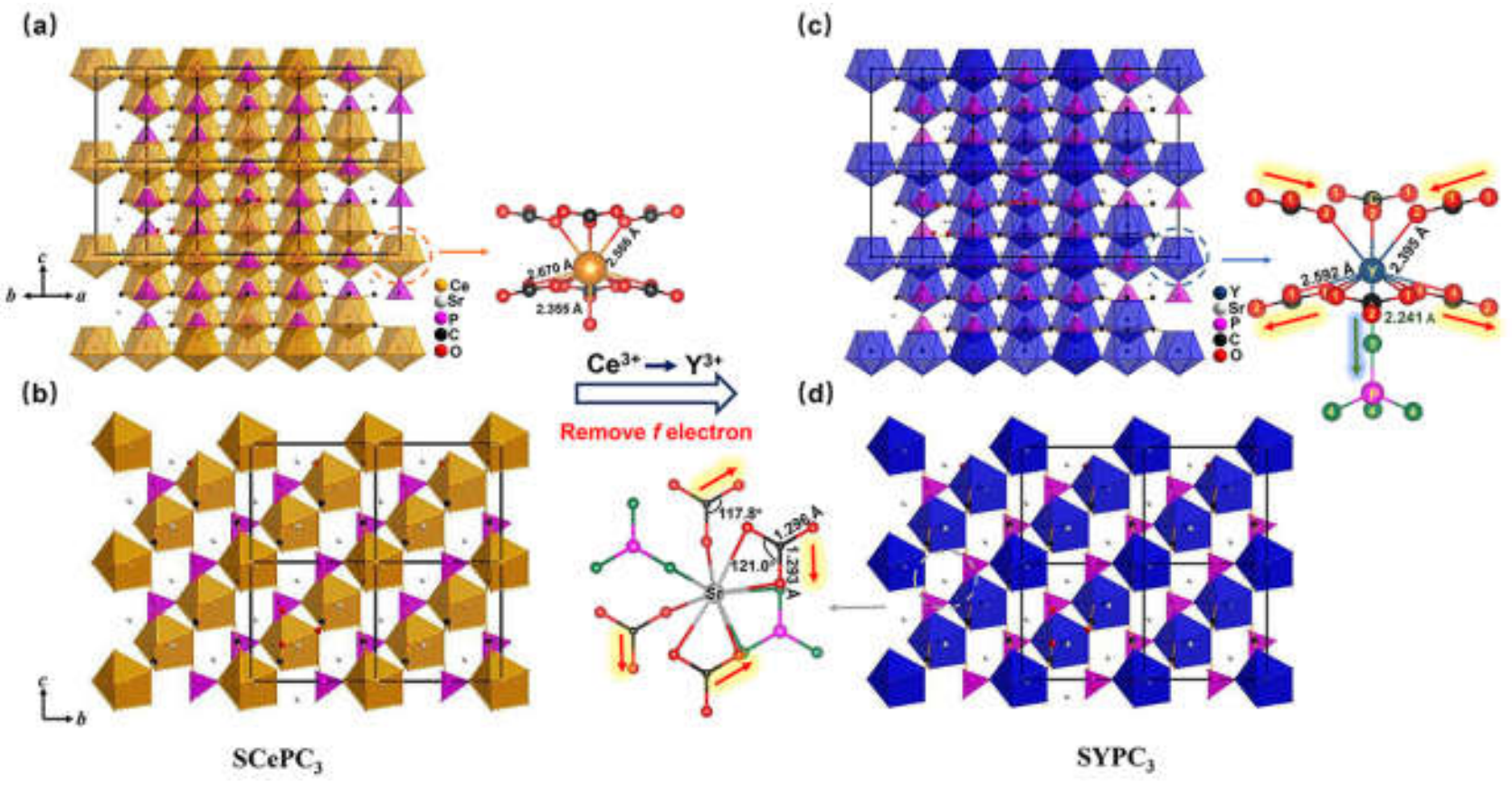

**Figure S1.** Crystal structure evolution from $Sr_3Ce[PO_4][CO_3]_3$ **$SCePC_3$** (a, b), to $Sr_3Y[PO_4][CO_3]_3$ **$SYPC_3$** (c, d). The polyhedron: orange-$CeO_{10}$ unit, violet-$YO_{10}$ unit, gray-$SrO_9$ unit, pink-$PO_4$ unit.

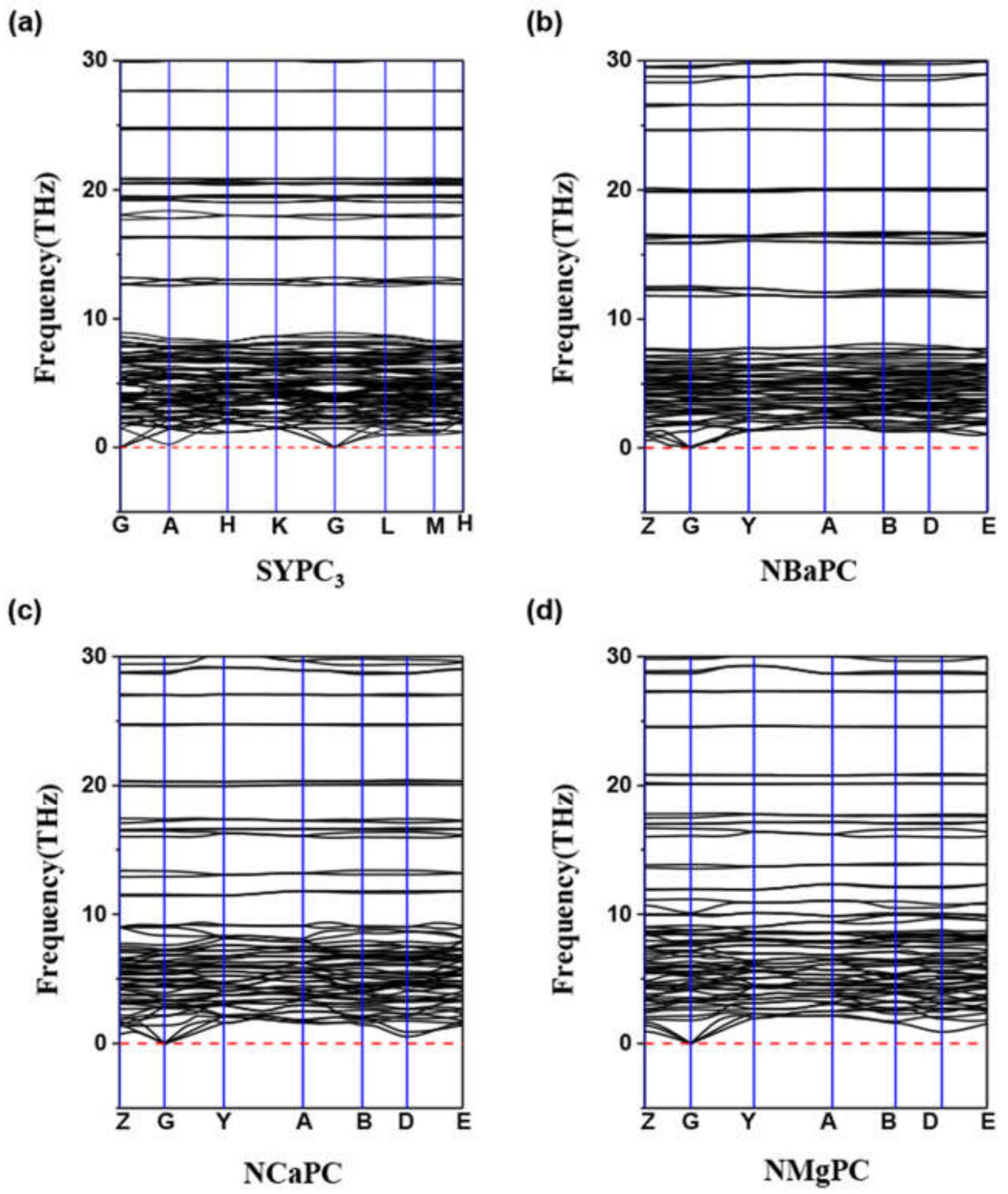

**Figure S2**. The phonon spectra of (a) $Sr_3Y[PO_4][CO_3]_3$, **$SYPC_3$**, and $Na_3X[PO_4][CO_3]$ (**NXPC,** X**=** (b) Ba, (c) Ca, (d) Mg). The calculated phonon vibrational spectra show that the imaginary phonon mode does not exist, which indicates that these four compounds are kinetically stable.

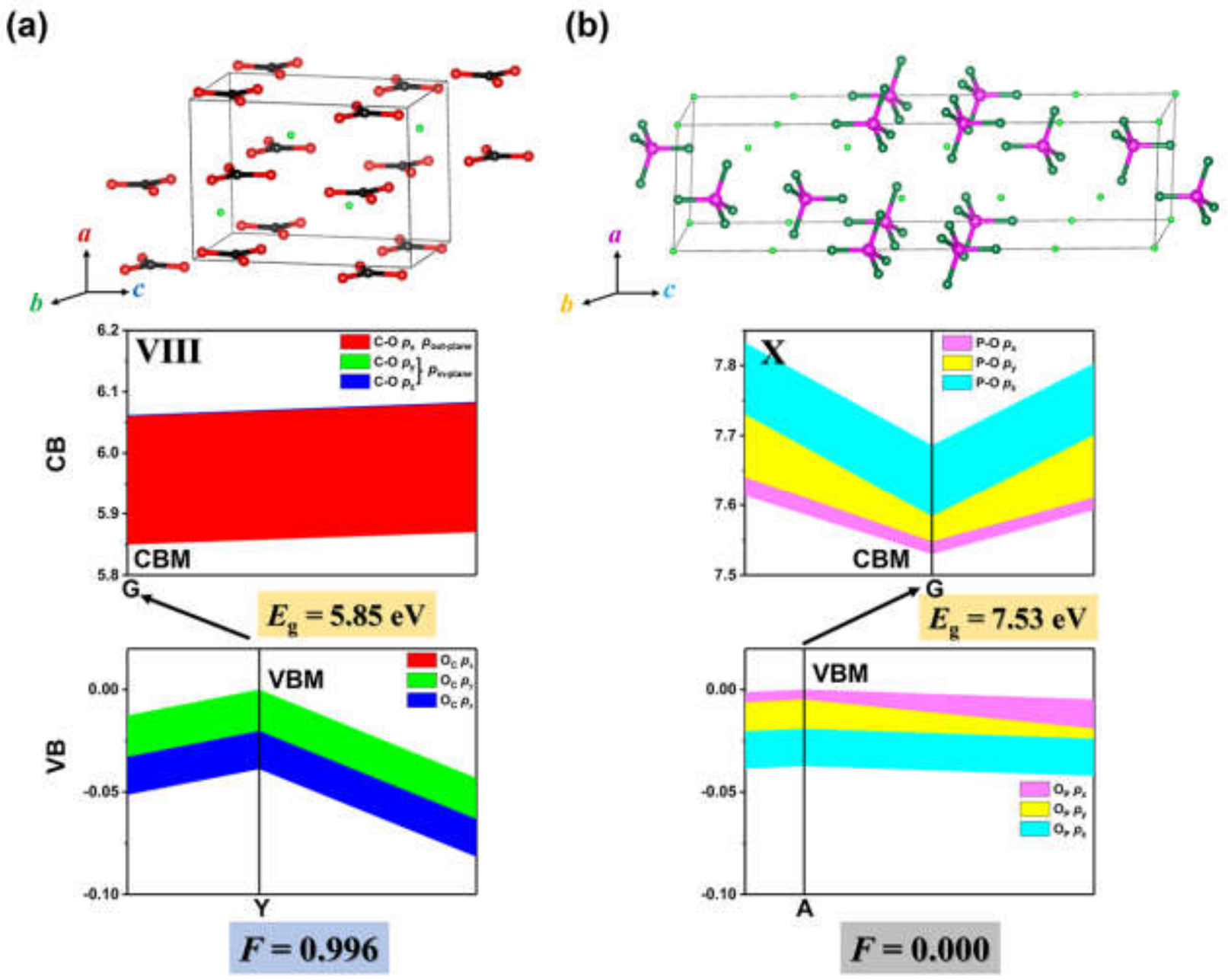

**Figure S3**. The crystal structure and orbital projected CBM and VBM along the $k_z$ direction in the Brillouin zone of (a) $Sr[CO_3]$ (**VIII**) and (b) $Sr_3[PO_4]_2$ (**X**). The $p_x$, $p_y$ and $p_z$ orbitals are projected on the VBM and the CBM, in which the curve width represents the orbital volume density (1/Å$^3$).

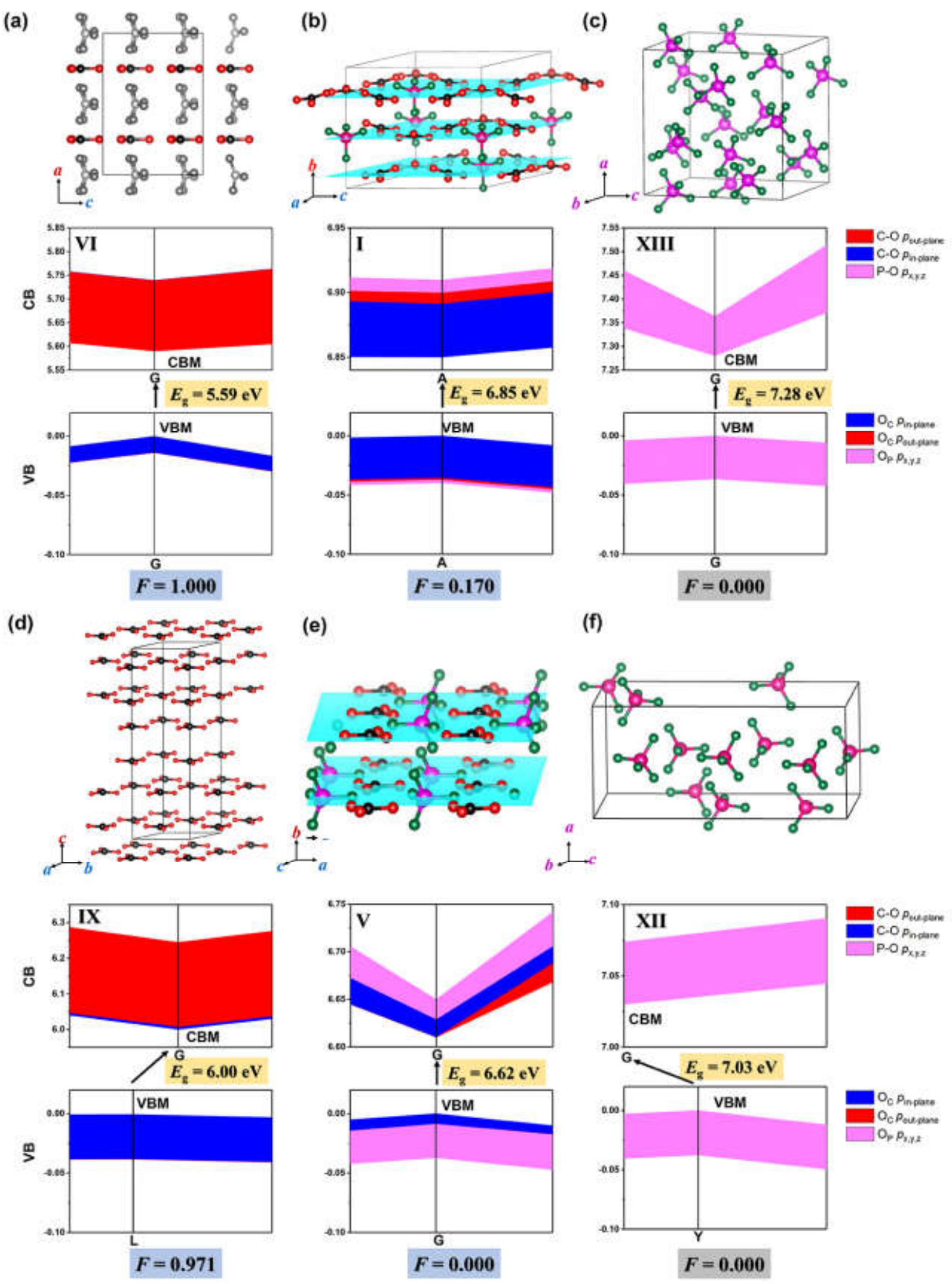

**Figure S4**. The crystal structure (with cations omitted) and orbital projected CBM and VBM along the $k_z$ direction in the Brillouin zone of (a) $Na_3Y(CO_3)_3$ (**VI**), (b) **SYPC$_3$** (**I**), (c) $Sr_3Y[PO_4]_3$ (**XIII**), (d) $NaMg[CO_3]_2$ (**IX**), (e) **NMgPC** (**V**), (f) $NaMgPO_4$ (**XII**), respectively. The *p* orbitals of $[CO_3]^{2-}$ or $[PO_4]^{3-}$ are projected on the VBM and the CBM, in which the curve width represents the orbital volume density (1/Å$^3$).

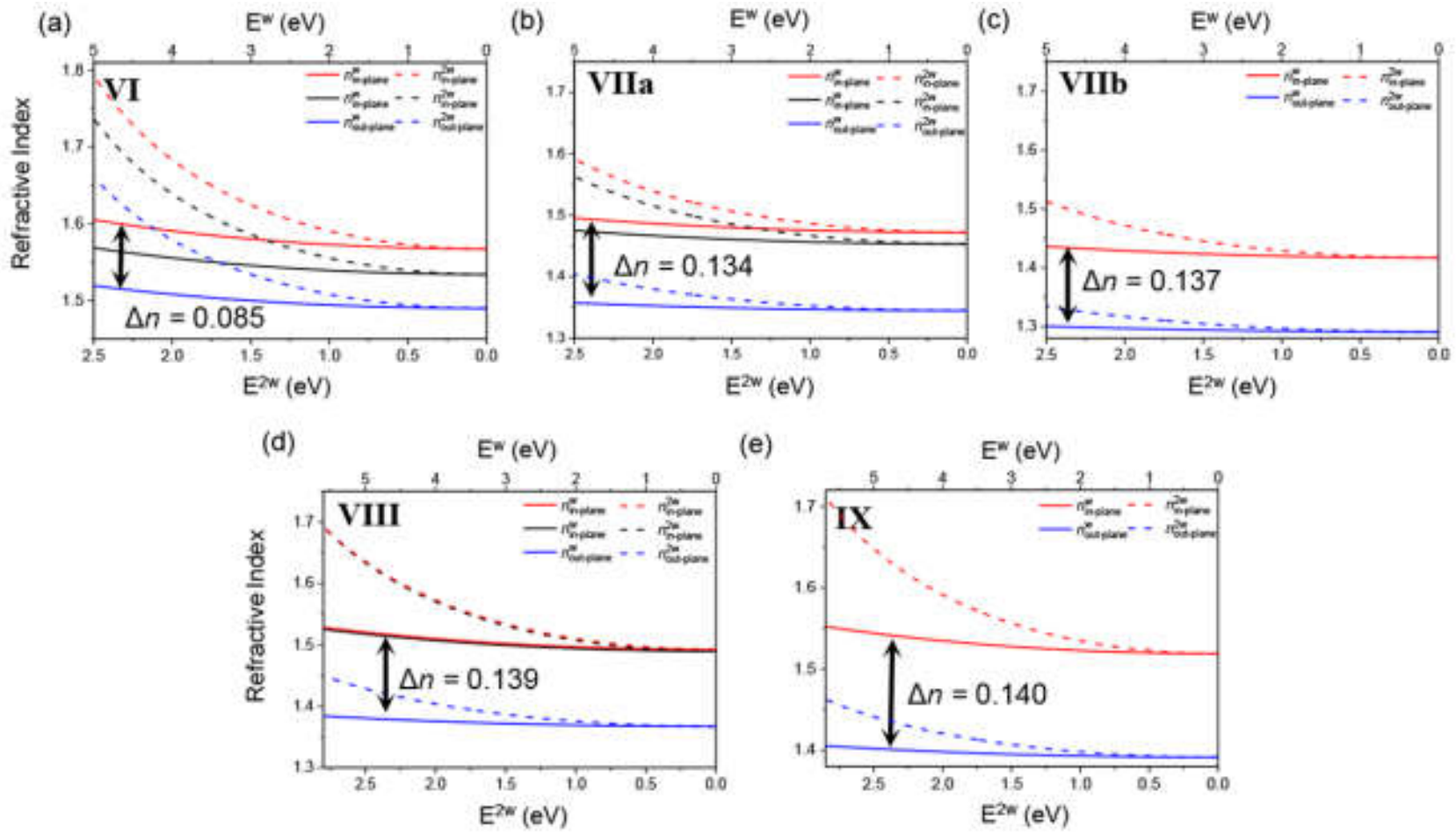

**Figure S5**. The calculated dispersion of the refractive indices of (a) $Na_3Y[CO_3]_3$ (**VI**), (b) $Na_2CO_3$ (**VIIa**), (c) $Na_2CO_3$ (**VIIb**), (d) $SrCO_3$ (**VIII**), (e) $Na_2Mg[CO_3]_2$ (**IX**).

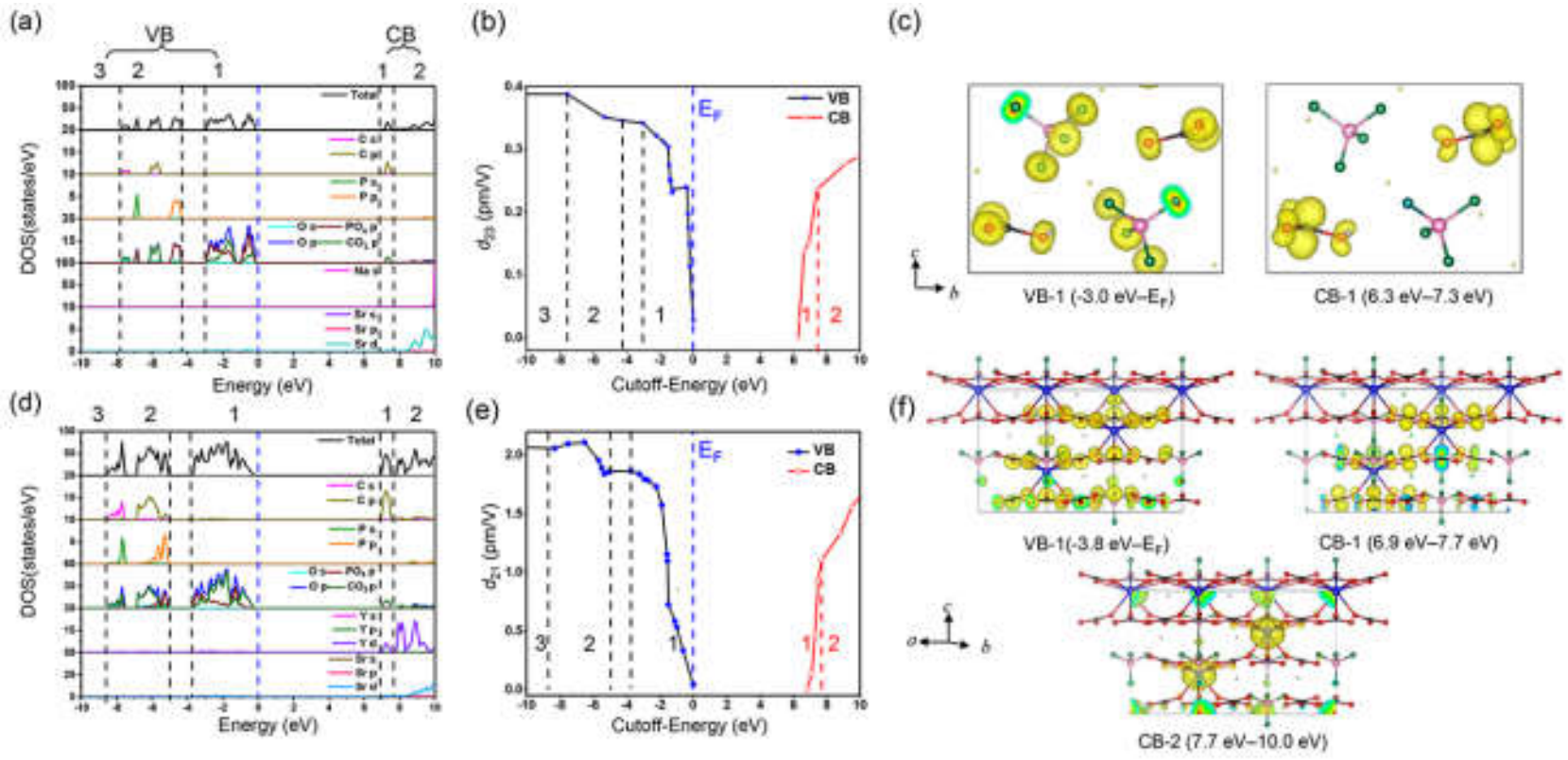

**Figure S6**. Calculated results of **NSrPC** and **$SYPC_3$**: (a), (d) total and partial DOS; (b), (e) cut-off energy dependences of the max static $d_{ij}$ coefficient (c) charge density maps of **NSrPC** in the VB-1, CB-1 alone the *a* axis. (f) charge density maps of **$SYPC_3$** in the VB-1, CB-1 and CB-2 alone the *b* axis.

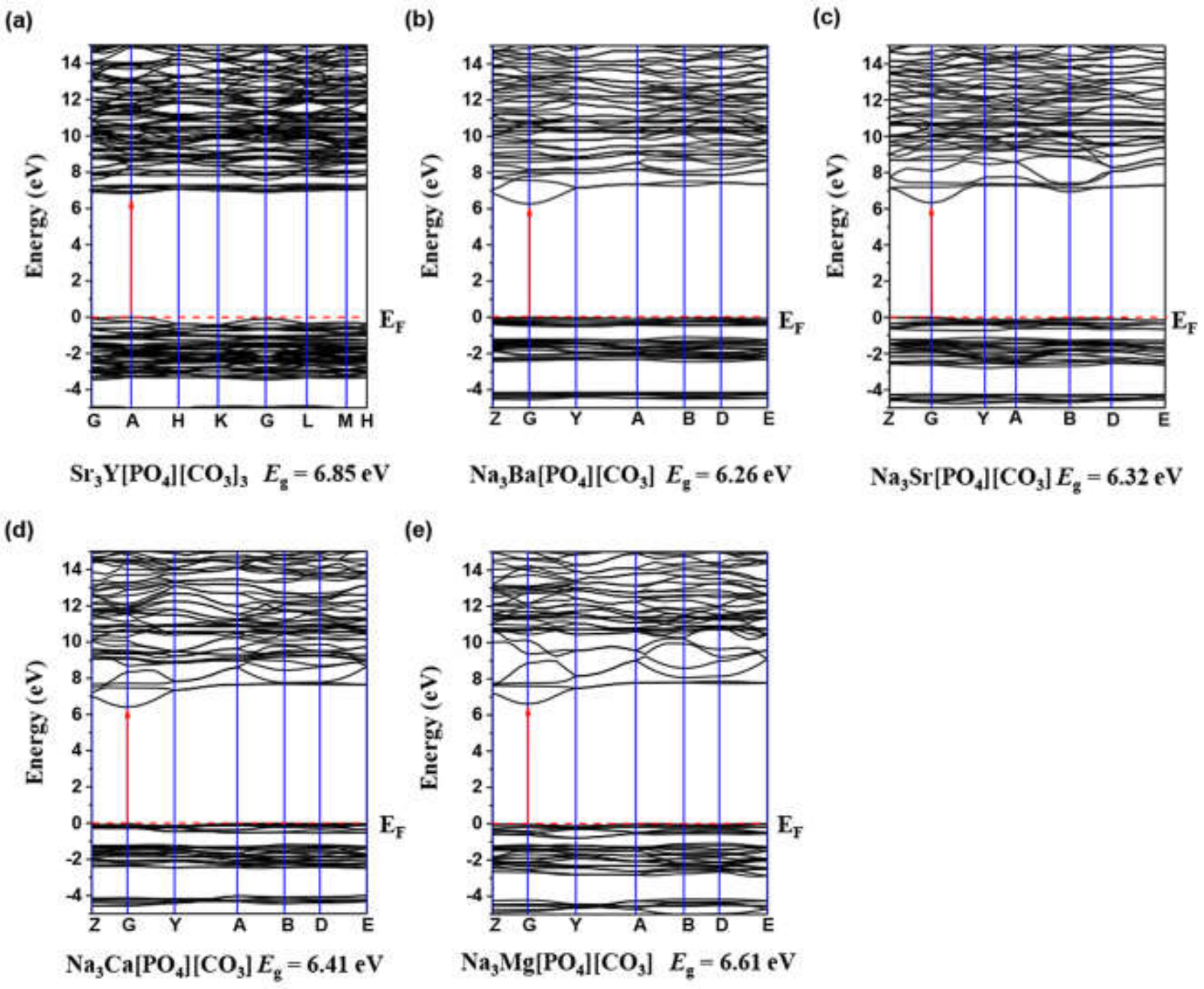

**Figure S7.** Band structures of (a) **SYPC$_3$**, (b) **NBaPC**, (c) **NSrPC**, (d) **NCaPC**, (e) **NMgPC**.

**Supporting Information**
**Tables**

**Table S1.** List of all the known noncentrosymmetric carbonophosphates, three of them are found as a mineral.

| No | Compounds | Crystal class/ Space Group | Cation | Anion | Mineral | Ref. |
|---|---|---|---|---|---|---|
| 1 | $Sr_3Ce[PO_4][CO_3]_3$ | 3*m*/*R*3*m* | $Sr^{2+}$, $Ce^{3+}$ | $CO_3^{2-}$, $PO_4^{3-}$ | Daqingshanite | 5 |
| 2 | $Na_3Sr[PO_4][CO_3]$ **(NSrPC)** | 2/*P*2$_1$ | $Na^+$, $Sr^{2+}$ | $CO_3^{2-}$, $PO_4^{3-}$ | Crawfordite | 6 |
| 3 | $Ca_{9.75}[PO_4]_{5.5}[CO_3]_{1.5}$ | $\bar{6}/P\bar{6}$ | $Ca^{2+}$ | $CO_3^{2-}$, $PO_4^{3-}$ | A-type Carbonate apatite | 7 |
| 4 | $Ca_{9.9}[PO_4]_6[CO_3]_{0.9}$ | *m*/*Pc* | | | A-type Carbonate apatite | 8 |
| 5 | $Na_{1.2}Ca_{8.8}[PO_4]_{4.8}[CO_3]_{2.2}$ | $\bar{6}/P\bar{6}$ | $Na^+$, $Ca^{2+}$ | $CO_3^{2-}$, $PO_4^{3-}$ | AB-type Carbonate apatite | 9 |

**Table S2.** Crystallographic data and structure optimizations of **SYPC$_3$** and **NXPC** (X = Ba, Ca, Mg). The only two sets of available experimental data are listed, which agree well with our optimization data indicating our calculation accuracy.

| | Space group | | *a*(Å) | *b*(Å) | *c*(Å) | *V*(Å$^3$) | Reference |
|---|---|---|---|---|---|---|---|
| $Sr_3Ce[PO_4][CO_3]_3$ | *R*3*m* | Exp. | 10.0730 | 10.0730 | 9.2340 | 811.41 | 5 |
| | | Cal. | 10.0331 | 10.0331 | 9.1507 | 797.73 | *This work* |
| | | Error | -0.4% | -0.4% | -0.9% | -1.7% | |
| $Sr_3Ce[PO_4][CO_3]_3$ **SYPC$_3$** | *R*3*m* | Cal. | 9.9495 | 9.9495 | 8.9391 | 766.35 | *This work* |
| | | | | | | | |
| $Na_3Ba[PO_4][CO_3]$ **NBaPC** | *P*2$_1$ | Cal. | 5.5837 | 7.2475 | 9.3191 | 376.70 | *This work* |
| | | | | | | | |
| $Na_3Sr[PO_4][CO_3]$ **NSrPC** | *P*2$_1$ | Exp. | 5.3831 | 7.08961 | 9.2836 | 354.30 | 6 |
| | | Cal. | 5.3358 | 7.1854 | 9.2096 | 352.75 | *This work* |
| | | Error | -0.9% | 1.4% | -0.8% | -0.4% | |
| $Na_3Ca[PO_4][CO_3]$ **NCaPC** | *P*2$_1$ | Cal. | 5.2596 | 7.0164 | 9.3036 | 343.32 | *This work* |
| | | | | | | | |
| $Na_3Mg[PO_4][CO_3]$ **NMgPC** | *P*2$_1$ | Cal. | 5.2949 | 6.5912 | 8.9848 | 313.55 | *This work* |
| | | | | | | | |

## Supporting Information

**Table S2-1.** Additional structure optimization results of $Sr_3Y[PO_4][CO_3]_3$, **$SYPC_3$**.

| $Sr_3Y[PO_4][CO_3]_3$ **$SYPC_3$** | | | | | | | |
|---|---|---|---|---|---|---|---|
| Space group | *R*3*m* | | | | | | |
| Unit cell Parameters | *a* (Å) | *b* (Å) | *c* (Å) | *α* (°) | *β* (°) | *γ* (°) | *V* (Å$^3$) |
| | 9.9495 | 9.9495 | 8.9391 | 90 | 90 | 120 | 766.35 |
| Atomic fractional coordinates | *x/a* | | *y/b* | | *z/c* | | Wyckoff |
| Y | 0.00000 | | 0.00000 | | 0.99815 | | 3*a* |
| Sr | 0.18437 | | 0.81563 | | 0.29062 | | 9*b* |
| P | 0.00000 | | 0.00000 | | 0.41978 | | 3*a* |
| O | 0.25911 | | 0.01803 | | 0.07760 | | 18*c* |
| O | 0.00000 | | 0.00000 | | 0.24890 | | 3*a* |
| O | 0.76042 | | 0.23958 | | 0.13368 | | 9*b* |
| O | 0.41714 | | 0.58286 | | 0.14825 | | 9*b* |
| C | 0.50014 | | 0.49986 | | 0.43386 | | 9*b* |

**Table S2-2.** Additional structure optimization results of $Na_3X[PO_4][CO_3]$, (**NXPC**, X = Ba, Ca, Mg).

| $Na_3Ba[PO_4][CO_3]$ **NBaPC** | | | | | | | |
|---|---|---|---|---|---|---|---|
| Space group | *P*2$_1$ | | | | | | |
| Unit cell Parameters | *a* (Å) | *b* (Å) | *c* (Å) | *α* (°) | *β* (°) | *γ* (°) | *V* (Å$^3$) |
| | 5.5837 | 7.2475 | 9.3191 | 90 | 92.69 | 90 | 376.71 |
| Atomic fractional coordinates | *x/a* | | *y/b* | | *z/c* | | Wyckoff |
| Ba | 0.03991 | | 0.74743 | | 0.67339 | | 2*a* |
| Na | 0.96903 | | 0.06392 | | 0.98259 | | 2*a* |
| Na | 0.52406 | | -0.02713 | | 0.50026 | | 2*a* |
| Na | 0.53769 | | 0.80436 | | 0.14523 | | 2*a* |
| P | 0.04402 | | 0.77661 | | 0.30781 | | 2*a* |
| C | 0.44383 | | 0.23595 | | 0.16113 | | 2*a* |
| O | 0.11897 | | 0.89860 | | 0.18037 | | 2*a* |
| O | 0.09968 | | 0.10540 | | 0.74985 | | 2*a* |
| O | 0.12283 | | 0.38987 | | 0.59372 | | 2*a* |
| O | 0.27333 | | 0.72345 | | 0.40083 | | 2*a* |
| O | 0.25398 | | 0.29298 | | 0.08869 | | 2*a* |
| O | 0.37592 | | 0.66862 | | 0.90516 | | 2*a* |
| O | 0.45562 | | 0.24534 | | 0.30217 | | 2*a* |

**(continued) Table S2-2.** Additional structure optimization results of $Na_3X[PO_4][CO_3]$, (**NXPC**, X = Ba, Ca, Mg).

| $Na_3Ca[PO_4][CO_3]$ **NCaPC** | | | | | | | |
|---|---|---|---|---|---|---|---|
| Space group | $P2_1$ | | | | | | |
| Unit cell Parameters | *a* (Å) | *b* (Å) | *c* (Å) | *α* (°) | *β* (°) | *γ* (°) | *V* (Å$^3$) |
| | 5.2596 | 7.0164 | 9.3036 | 90 | 90.35 | 90 | 343.33 |
| Atomic fractional coordinates | *x/a* | | *y/b* | | *z/c* | | Wyckoff |
| Ca | 0.00367 | | 0.73657 | | 0.28989 | | 2*a* |
| Na | 0.99992 | | 0.48640 | | 0.00009 | | 2*a* |
| Na | 0.50009 | | 0.48675 | | 0.49983 | | 2*a* |
| Na | 0.55108 | | 0.23620 | | 0.16602 | | 2*a* |
| P | 0.01258 | | 0.23652 | | 0.33773 | | 2*a* |
| C | 0.50034 | | 0.23652 | | 0.85873 | | 2*a* |
| O | 0.93419 | | 0.41517 | | 0.24715 | | 2*a* |
| O | 0.06527 | | 0.55792 | | 0.75307 | | 2*a* |
| O | 0.13980 | | 0.73626 | | 0.51923 | | 2*a* |
| O | 0.30197 | | 0.23669 | | 0.36932 | | 2*a* |
| O | 0.29782 | | 0.23650 | | 0.93827 | | 2*a* |
| O | 0.27511 | | 0.73637 | | 0.07969 | | 2*a* |
| O | 0.48373 | | 0.23671 | | 0.71957 | | 2*a* |

| $Na_3Mg[PO_4][CO_3]$ **NMgPC** | | | | | | | |
|---|---|---|---|---|---|---|---|
| Space group | $P2_1$ | | | | | | |
| Unit cell Parameters | *a* (Å) | *b* (Å) | *c* (Å) | *α* (°) | *β* (°) | *γ* (°) | *V* (Å$^3$) |
| | 5.2949 | 6.5912 | 8.9848 | 90 | 89.42 | 90 | 313.55 |
| Atomic fractional coordinates | *x/a* | | *y/b* | | *z/c* | | Wyckoff |
| Mg | 0.03337 | | 0.26348 | | 0.29342 | | 2*a* |
| Na | 0.00003 | | 0.51355 | | 0.00004 | | 2*a* |
| Na | 0.43645 | | 0.76361 | | 0.16294 | | 2*a* |
| Na | 0.50015 | | 0.01346 | | 0.50004 | | 2*a* |
| P | 0.02062 | | 0.26347 | | 0.65514 | | 2*a* |
| C | 0.50228 | | 0.26348 | | 0.16066 | | 2*a* |
| O | 0.07301 | | 0.57412 | | 0.25432 | | 2*a* |
| O | 0.07279 | | 0.95285 | | 0.25426 | | 2*a* |
| O | 0.10608 | | 0.76354 | | 0.50044 | | 2*a* |
| O | 0.25746 | | 0.76355 | | 0.87668 | | 2*a* |
| O | 0.31034 | | 0.26340 | | 0.64027 | | 2*a* |
| O | 0.33109 | | 0.26350 | | 0.05704 | | 2*a* |
| O | 0.43322 | | 0.26338 | | 0.29867 | | 2*a* |

## Supporting Information

**Table S3.** Calculated elastic constant $C_{ij}$, bulk modulus B (GPa), Poisson ratio and elastic anisotropy index of **SYPC₃** and **NXPC** (X = Ba, Ca, Mg).

| | | | | | | | | | | |
|---|---|---|---|---|---|---|---|---|---|---|
| **Trigonal *R*3*m*** | | | | | | | | | | |
| **SYPC₃** | $C_{11}$ | $C_{12}$ | $C_{13}$ | $C_{14}$ | $C_{33}$ | $C_{44}$ | | B [a] | Poisson [a] | Elastic anisotropy [b] |
| | 135.16 | 72.47 | 68.67 | 10.59 | 155.58 | 43.51 | | 93.95 | 0.321 | 0.4954 |
| **Monoclinic *P*2₁** | | | | | | | | | | |
| **NBaPC** | $C_{11}$ | $C_{22}$ | $C_{33}$ | $C_{44}$ | $C_{55}$ | $C_{66}$ | | 43.94 | 0.280 | 0.3360 |
| | 82.14 | 71.48 | 83.45 | 17.80 | 18.99 | 23.69 | | | | |
| | $C_{12}$ | $C_{13}$ | $C_{15}$ | $C_{23}$ | $C_{25}$ | $C_{35}$ | $C_{46}$ | | | |
| | 25.99 | 29.28 | 6.50 | 23.95 | -5.79 | 2.27 | -0.30 | | | |
| | | | | | | | | | | |
| **NCaPC** | $C_{11}$ | $C_{22}$ | $C_{33}$ | $C_{44}$ | $C_{55}$ | $C_{66}$ | | 56.62 | 0.276 | 1.2716 |
| | 113.66 | 117.10 | 100.69 | 11.54 | 31.00 | 25.69 | | | | |
| | $C_{12}$ | $C_{13}$ | $C_{15}$ | $C_{23}$ | $C_{25}$ | $C_{35}$ | $C_{46}$ | | | |
| | 23.18 | 40.84 | 0.88 | 25.07 | 5.51 | 5.79 | -0.23 | | | |
| | | | | | | | | | | |
| **NMgPC** | $C_{11}$ | $C_{22}$ | $C_{33}$ | $C_{44}$ | $C_{55}$ | $C_{66}$ | | 61.53 | 0.281 | 0.6193 |
| | 119.98 | 114.89 | 113.68 | 16.95 | 29.18 | 29.50 | | | | |
| | $C_{12}$ | $C_{13}$ | $C_{15}$ | $C_{23}$ | $C_{25}$ | $C_{35}$ | $C_{46}$ | | | |
| | 30.63 | 35.82 | 5.26 | 36.18 | 5.65 | 9.58 | 0.43 | | | |

[a] The bulk moduli and Poisson ratio used Voigt approximate; [b] Elastic Anisotropy defined as Universal Elastic Anisotropy Index ((Au=5*Gv/Gr+Kv/Kr-6))

(i) For **SYPC₃**, the criteria for mechanical stability of trigonal symmetry group are given by: $C_{ij}>0$, i=j=1~6; $C_{11}-C_{12}>0$; $C_{11}C_{33}C_{44}-C_{14}^2C_{33}-C_{13}^2C_{44}>0$; $C_{11}C_{33}+C_{12}C_{33}-2C_{13}^2>0$; $C_{44}(C_{11}-C_{12})-2C_{14}^2>0$. So **SYPC₃** is mechanically stable under ambient conditions.

(ii) For **NXPC** (X = Ba, Ca, Mg), the criteria for mechanical stability of monoclinic symmetry group are given by: $C_{ij}>0$, i=j=1~6; $C_{11}+C_{22}+C_{33}+2C_{12}+2C_{13}+2C_{23}>0$; $C_{33}C_{55}-C_{35}^2>0$; $C_{44}C_{66}-C_{46}^2>0$; $C_{22}+C_{33}-2C_{23}>0$; $C_{22}(C_{33}C_{55}-C_{35}^2)+2C_{23}C_{25}C_{35}-C_{23}^2C_{55}-C_{25}^2C_{33}>0$; $2C_{15}C_{25}(C_{33}C_{12}-C_{13}C_{23})+2C_{15}C_{35}(C_{22}C_{13}-C_{12}C_{23})+2C_{25}C_{35}(C_{11}C_{23}-C_{12}C_{13})-C_{15}^2(C_{22}C_{33}-C_{23}^2)-C_{25}^2(C_{11}C_{33}-C_{13}^2)-C_{35}^2(C_{11}C_{22}-C_{12}^2)+C_{55}C_{11}C_{22}C_{33}-C_{11}C_{23}^2-C_{22}C_{13}^2-C_{33}C_{12}^2+2C_{12}C_{13}C_{23}>0$. So, **NXPC** are mechanically stable under ambient conditions.

## Supporting Information

**Table S4.** The reliability of our calculation method which is confirmed by the testing calculations on the benchmark $KH_2PO_4$ **(KDP)** and other selected carbonates and phosphates.

| Major feature | Comp. | Experimental obv. | | | Calculated in this work | | | | Ref. |
|---|---|---|---|---|---|---|---|---|---|
| | | $E_g$ | SHG (× KDP) | $\Delta n$ [a] | $E_g$ [b] | $d_{ij}$ (pm/V) | $\Delta n$ [c] | $\lambda_{PM}$ (nm) [d] | |
| **Phosphates** | | | | | | | | | |
| $[H_2PO_4]^-$ | $KH_2PO_4$ (KDP) | 7.00 | 1 | 0.042 | 7.03 | 0.76 | 0.041 | 256 | 10 |
| $PO_4$ monomer | $LiCs_2PO_4$ | 7.02 | 2.6 | / | 7.05 | 2.88 | 0.011 | 502 | 11 |
| $P_2O_7$ dimer | $K_4Mg_4(P_2O_7)_3$ | 7.29 | 1.3 | / | 7.21 | 1.50 | 0.009 | 494 | 12 |
| $P_3O_{10}$ trimer | $Ba_3P_3O_{10}Cl$ | 6.89 | 0.6 | / | 6.95 | 0.53 | 0.029 | 474 | 13 |
| $[PO_3]_\infty$ chain | $KBa_2(PO_3)_5$ | 7.43 | 0.9 | / | 7.45 | 0.98 | 0.010 | 475 | 14 |
| **Carbonates** | | | | | | | | | |
| Fluorine | $KSrCO_3F$ | 6.36 | 3.3 | 0.111 | 6.63 | 2.56 | 0.110 | 201 | 15 |
| cation with lone-pair electrons | $CsPbCO_3F$ | 4.15 | 13.0 | / | 4.15 | 10.24 | 0.169 | 300 | 16 |
| rare earth elements | $Na_5Sc(CO_3)_4{\cdot}2H_2O$ | 5.59 | 1.8 | / | 5.70 | 1.21 | 0.035 | 367 | 17 |
| | $Na_4La_2(CO_3)_5$ | 5.29 | 3.0 | / | 5.41 | 2.28 | 0.044 | 423 | 18 |
| | $Y_8O(CO_3)_3(OH)_{15}Cl$ | 5.30 | 2.5 | 0.045 | 5.26 | 1.51 | 0.044 | 434 | 19 |
| cation with lone-pair electrons | $Bi_2O_2CO_3$ | 3.42 | 5.0 | / | 3.50 | 3.97 | 0.127 | 363 | 20 |

[a] characterized by using the polarizing microscope equipped (ZEISS Axio Scope. A1) with Berek compensator. The wavelength of the light source was 546.1 nm. [b] calculated by using the HSE06 exchange-correlation functional for accuracy. [c] calculated at 546.1 nm by using a scissor operator. [d] indicated by the calculated dispersion of the refractive indices.

## Supporting Information

**Table S5.** The absorption edge ($\lambda_{cutoff}$), birefringence value ($\Delta n$), shortest SHG phase-matching wavelength ($\lambda_{PM}$) and SHG effect of borate, carbonate, phosphate-based NLO compounds.

| | **Crystal** | **Structure units** | | **$\lambda_{cutoff}$ (nm)** | **$\Delta n$** | **$\lambda_{PM}$ (nm)** | **SHG effects ( × KDP)** | **Ref.** |
|---|---|---|---|---|---|---|---|---|
| | | *π*-conjugated | *Non-π*-conjugated | | | | | |
| | Borates | | | | | | | |
| 1 | $KB_5O_8·4H_2O$ | $[B_3O_7]^{5-}$ | / | 191 | 0.065 @ 546.1 nm | > 200 | 0.1 | 21 |
| 2 | *β*-$BaB_2O_4$ | $[B_3O_6]^{3-}$ | / | 190 | 0.114 @ 1064 nm | 205 | 3.0 | 21 |
| 3 | $LiSr(BO_3)_2$ | $[BO_3]^{3-}$ | / | 186 | 0.056 @ 1064 nm[a] | > 200 | 2.0 | 22 |
| 4 | $YCa_4O(BO_3)_3$ | $[BO_3]^{3-}$ | / | 202 | 0.045 @ 546.1 nm | 362 | 3.0 | 21 |
| 5 | $GdCa_4O(BO_3)_3$ | $[BO_3]^{3-}$ | / | 310 | 0.035 @ 546.1 nm | > 200 | 4.3 | 21 |
| 6 | $LiB_3O_5$ (LBO) | $[B_3O_7]^{5-}$ | / | 160 | 0.042 @ 589 nm | 277 | 3.0 | 23 |
| 7 | $Li_2B_4O_7$ (LB4) | $[B_3O_7]^{5-}$ | / | 160 | 0.052 @ 1064 nm | 244 | 0.4 | 21 |
| 8 | $CsB_3O_5$ (CBO) | $[B_3O_7]^{5-}$ | / | 167 | 0.050 @ 1064 nm | > 200 | 2.6 | 21 |
| 9 | $CsLiB_6O_{10}$ (CLBO) | $[B_3O_7]^{5-}$ | / | 180 | 0.049 @ 1064 nm | 243 | 2.2 | 24 |
| 10 | $Sr_2Be_2B_2O_7$ | $[BO_3]^{3-}$ | $[BeO_4]^{6-}$ | 155 | 0.062 @ 589 nm | 200 | 3.8 | 25 |
| 11 | $KBe_2BO_3F_2$ (KBBF) | $[BO_3]^{3-}$ | $[BeO_3F]^{5-}$ | 147 | 0.077 @ 1064 nm | 161 | 1.3 | 26 |
| 12 | $CsBe_2BO_3F_2$ (CBBF) | $[BO_3]^{3-}$ | $[BeO_3F]^{5-}$ | 151 | 0.058 @ 1064 nm | > 200 | 1.2 | 27 |
| 13 | $NaBe_2BO_3F_2$ (NBBF) | $[BO_3]^{3-}$ | $[BeO_3F]^{5-}$ | 155 | 0.091 @ 200 nm[a] | 185 | 1.4 | 28 |
| 14 | $RbBe_2BO_3F_2$ (RBBF) | $[BO_3]^{3-}$ | $[BeO_3F]^{5-}$ | 152 | 0.073 @ 1064 nm | 170 | 1.1 | 29 |
| 15 | $NH_4Be_2BO_3F_2$ (ABBF) | $[BO_3]^{3-}$ | $[BeO_3F]^{5-}$ | 153 | 0.057 @ 400 nm | 174 | 1.2 | 30 |
| 16 | $Be_2BO_3F$ | $[BO_3]^{3-}$ | $[BeO_3F]^{5-}$ | 150 | 0.091 @ 1064 nm[a] | 180 | 0.1 | 31 |
| 17 | $BaBe_2BO_3F_3$ | $[BO_3]^{3-}$ | $[BeO_3F]^{5-}$ | 182 | 0.081 @ 200 nm[a] | 196 | 0.1 | 32 |
| 18 | $K_3Ba_3Li_2Al_4B_6O_{10}F$ | $[BO_3]^{3-}$ | $[BeO_4]^{6-}$, $[AlO_4]^{5-}$ | 190 | 0.052 @ 532 nm | > 200 | 1.5 | 33 |
| 19 | $CsAlB_3O_6F$ | $[B_3O_6]^{3-}$ | $[AlO_3F]^{4-}$ | 181 | 0.091 @ 1064 nm[a] | 182 | 2.0 | 34 |
| 20 | $NH_4B_4O_6F$ | $[BO_3]^{3-}$ | $[BO_3F]^{4-}$ | 156 | 0.117 @ 1064 nm | 158 | 3.0 | 35 |
| 21 | $CsB_4O_6F$ | $[B_3O_6]^{3-}$ | $[BO_3F]^{4-}$ | 155 | 0.114 @ 1064 nm | 172 | 1.9 | 36 |
| 22 | $CaB_5O_7F_3$ | $[BO_3]^{3-}$ | $[BO_3F]^{4-}$ | 178 | 0.070 @ 1064 nm | 183 | 2.0 | 37 |
| | Carbonates | | | | | | | |
| 23 | $(NH_4)_2Ca_2Y_4(CO_3)_9·H_2O$ | $[CO_3]^{2-}$ | / | 190 | 0.040 @ 1064 nm[a] | 260 | 2.1 | 38 |
| 24 | $Y_8O(OH)_{15}(CO_3)_3Cl$ | $[CO_3]^{2-}$ | / | 234 | 0.045 @ 1064 nm[a] | > 200 | 2.5 | 19 |
| 25 | $Na_2Ca_2(CO_3)_3$ | $[CO_3]^{2-}$ | / | 205 | 0.061 @ 1064 nm[a] | > 200 | 3.0 | 39 |
| 26 | $Na_6Ca_5(CO_3)_8$ | $[CO_3]^{2-}$ | / | 200 | 0.082 @ 1064 nm[a] | > 200 | 1.0 | 39 |
| 27 | $CsNa_5Ca_5(CO_3)_8$ | $[CO_3]^{2-}$ | / | 209 | 0.063 @ 1064 nm[a] | > 200 | 1.0 | 18 |
| 28 | $Na_4La_2(CO_3)_5$ | $[CO_3]^{2-}$ | / | 235 | 0.043 @ 1064 nm[a] | 423 | 3.0 | 18 |
| 29 | $Na_3Y(CO_3)_3$ | $[CO_3]^{2-}$ | / | 220 | 0.085 @ 1064 nm[a] | 258 | 4.6 | 40 |
| 30 | $KSrCO_3F$ | $[CO_3]^{2-}$ | $F^-$ | 195 | 0.105 @ 1064 nm | 200 | 3.3 | 15 |
| 31 | $RbSrCO_3F$ | $[CO_3]^{2-}$ | $F^-$ | 197 | 0.102 @ 1064 nm[a] | 197 | 3.3 | 41 |
| 32 | $KCaCO_3F$ | $[CO_3]^{2-}$ | $F^-$ | 197 | 0.112 @ 1064 nm[a] | 197 | 3.6 | 41 |
| 33 | $CsCaCO_3F$ | $[CO_3]^{2-}$ | $F^-$ | 198 | 0.107 @ 1064 nm[a] | 198 | 1.1 | 41 |
| Nitrates | | | | | | | | |
| 34 | $Ba_2NO_3(OH)_3$ | $[NO_3]^-$ | $[OH]^-$ | 190 | 0.080 @ 532 nm | < 200 | 4.0 | 42 |
| 35 | $Sr_2NO_3(OH)_3$ | $[NO_3]^-$ | $[OH]^-$ | 194 | 0.076 @ 532 nm | < 200 | 3.6 | 43 |
| 36 | $Rb_2Na(NO_3)_3$ | $[NO_3]^-$ | / | 270 | 0.091 @ 532 nm[a] | > 200 | 5.0 | 44 |
| Phosphates | | | | | | | | |
| 37 | $KH_2PO_4$ | / | $[H_2PO_4]^-$ | 170 | 0.041 @ 1064 nm | 256 | 1.0 | 10 |
| 38 | $NH_4H_2PO_4$ | / | $[H_2PO_4]^-$ | 184 | 0.034 @1064 nm | 258 | 1.2 | 10 |
| 39 | $KLa(PO_3)_4$ | / | $[PO_3]_\infty^-$ | 162 | 0.008 @ 1064 nm[a] | 510 | 0.7 | 45 |
| 40 | $KBa_2(PO_3)_5$ | / | $[PO_3]_\infty^-$ | 167 | 0.014 @ 1064 nm[a] | 475 | 0.9 | 14 |
| 41 | $RbBa_2(PO_3)_5$ | / | $[PO_3]_\infty^-$ | 163 | 0.016 @ 1064 nm[a] | 460 | 1.4 | 46 |
| 42 | $BaP_3O_{10}Cl$ | / | $[P_3O_{10}]^{5-}$ | 180 | 0.029 @ 1064 nm[a] | 474 | 0.6 | 13 |
| 43 | $K_4Mg_4(P_2O_7)_3$ | / | $[P_2O_7]^{4-}$ | 170 | 0.011 @ 1064 nm | 494 | 1.3 | 12 |
| 44 | $LiCs_2PO_4$ | / | $[PO_4]^{3-}$ | 176 | 0.010 @ 1064 nm[a] | 502 | 2.6 | 11 |
| 45 | $NaNH_4PO_3F\cdot H_2O$ | | $[PO_3F]^{2-}$ | 176 | 0.053 @ 589.1 nm | 194 | 1.0 | 47 |

[a] calculated by using the first principles calculation method. The serial numbers is listed to represent each compound in **Figure 9**

# Supporting Information